\documentclass[useAMS,usenatbib]{mn2e}
\usepackage{amssymb}
\usepackage{graphicx}
\def\hi{H\,{\sc i}}
 
\def\deg{$^{\circ}$}
\def\kms{km~s$^{-1}$}
\def\msun{$M_{\odot}$}

\def\aj{AJ}%
\def\araa{ARA\&A}%
\def\apj{ApJ}%
\def\apjl{ApJ}%
\def\apjs{ApJS}%
\def\aap{A\&A}%
\def\mnras{MNRAS}%
\def\nat{Nature}%
\title[The dark matter content of the blue compact dwarf NGC 2915]{The dark matter content of the blue compact dwarf\\ NGC 2915}
\author[Elson et al.]{E. C. Elson$^{1}$\thanks{E-mail:
elson.e.c@gmail.com, edeblok@ast.uct.ac.za, kraan@ast.uct.ac.za}, W. J. G. de Blok$^{1}$ and R. C. Kraan-Korteweg$^{1}$\\
$^{1}$Department of Astronomy, University of Cape Town, Private Bag X3, Rondebosch 7701 , South Africa}
\begin{document}

\date{31 January 2010}

%\pagerange{\pageref{firstpage}--\pageref{lastpage}} \pubyear{2010}
\pagerange{000--000} \pubyear{2010}

\maketitle

\label{firstpage}

\begin{abstract}
NGC 2915 is a nearby blue compact dwarf with the \hi\ properties of a late-type spiral.  Its large, rotating \hi\ disk (extending out to $R\sim 22$~$B$-band scale lengths) and apparent lack of stars in the outer \hi\ disk make it a useful candidate for dark matter studies.  New \hi\ synthesis observations of NGC 2915 have been obtained using the Australian Telescope Compact Array.  These data are combined with high-quality 3.6~\micron\ imaging from the \emph{Spitzer} Infrared Nearby Galaxies Survey.  The central regions of the \hi\ disk are shown to consist of two distinct \hi\ concentrations with significantly non-Gaussian line profiles.  We fit a tilted ring model to the \hi\ velocity field to derive a rotation curve.  This is used as input for mass models that determine the contributions from the stellar and gas disks as well as the dark matter halo.  The galaxy is dark-matter-dominated at nearly all radii.  At the last measured point of the rotation curve, the total mass to blue light ratio is $M_{tot}/L_B\sim$ 140 \msun/L$_B$, making NGC~2915 one of the darkest galaxies known.  We show  that the stellar disk cannot account for the steeply-rising portion of the observed rotation curve.  The best-fitting dark matter halo is a pseudo-isothermal sphere with a core density $\rho_0\sim 0.17 \pm 0.03$ \msun~pc$^{-3}$ and a core radius $r_c\sim 0.9 \pm 0.1$~kpc. 
\end{abstract}


\section{Introduction}
Dwarf galaxies are thought to dominate the cosmic scenery in terms of number density \citep[][]{springel_2005,sandage_virgo_LF,mateo_dwarfs_review_1998} and therefore form an important morphological class of galaxies.  Several investigators have studied the mass distributions of these galaxies \citep[e.g.][and references therein]{carignan_beaulieu_DDO154_1989,cote_carignan_sancisi_1991,broeils_phd,meurer2,deblok_mcgaugh_1996}.  \citet{swaters_thesis} was the first to study the dark matter (DM) properties of a large representative sample of nearby dwarf galaxies as part of the Westerbork \hi\ Survey of Spiral and Irregular Galaxies \citep[WHISP, ][]{WHISP_swaters}.  More recently, a morphologically diverse sample of nearby galaxies, including dwarf systems, was observed as part of The~\hi\ Nearby~Galaxy~Survey \citep[THINGS, ][]{THINGS_walter}.  These cumulative efforts have led to an understanding that DM plays an important role in the dynamics and evolution of dwarf galaxies.  

\citet{swaters_thesis} found that none of the observed rotation curves in his sample of late-type dwarf galaxies decline at outer radii.  This was further confirmation that these systems have a large dark mass fraction.  Dwarf galaxies therefore serve as ideal candidates to test theories of DM.  Observational studies of the mass distributions of these dwarf systems have shown their DM halos to be well-modelled by a pseudo iso-thermal sphere with an approximately constant-density core \citep{deblok_et_al_1996,cote_carignan_freeman_2000,THINGS_deblok}.  Theoretically, the study of the properties of DM halos in a hierarchical clustering context is carried out mainly in the form of Cold Dark Matter (CDM) numerical simulations \citep[e.g.][]{NFW,moore_1999,aquarius_subhalos}.   The equilibrium density profile, $\rho(r)$, of simulated DM halos is found to vary with radius, $r$, as $\rho(r) \propto r^{\alpha}$ with $\alpha \approx -1$ \citep{NFW_1997}, thereby predicting extremely steep inner density profiles.  This discrepancy between the observed shapes of DM halos and the predictions from numerical simulations has become known as the ``cusp/core'' problem. 

%Proponents of the results of the CDM simulations argue that various systematic effects such as beam smearing and pointing offsets can make the intrinsically cuspy inner density profile of a galaxy appear core-like \citep[][]{}.  Non-circular gas motions can also lead to erroneous conclusions from dynamical analyses.  Observers argue, however, that the non-circular motions would have to be large ($\sim 20$ \kms) and cover a large portion of a galaxy's disk for an intrinsically cuspy core to be measured as a shallow, constant density core \citep{deblok_bosma_mcgaugh_2003}.  Non-circular motions measured using 2-dimensional velocity fields are typically a few \kms\ \citep{trachternach_THINGS}.  %Despite the seemingly contradicting results from numerical simulations and observations, it is clear that further detailed studies of late-type dwarf galaxies will help us to better understand the role of DM in these systems.%  

Attempts at reconciling observations with theoretical predictions depend crucially on accurate dynamical analyses of nearby galaxies.  To test DM halo models one wants galaxies that are as dark-matter-dominated as possible.  In this respect, a potentially useful candidate for DM studies is NGC~2915.  This galaxy is classified as a nearby \citep[$D\sim$~3.78~Mpc, ][]{karachentsev_catalog} blue compact dwarf according to its optical appearance, yet has the \hi\ morphology of a late-type spiral.  Detailed imaging by \citet{meurer1} showed that the optical appearance of NGC~2915 is dominated by two main stellar populations: a compact blue population, which is the location of on-going high-mass star formation; and a more diffuse, older red population.   What makes this galaxy so interesting is the fact that, when observed at 21~cm, the stellar core of NGC~2915 is seen to be completely embedded in a huge \hi\ disk extending out to $R\sim$~22 $B$-band scale lengths \citep{meurer2}.  Furthermore, this gas disk has well-defined spiral structure that is not seen in the stellar disk.  No significant star-formation is observed in the outer parts of this gas disk.  Only a few faint HII regions were detected by \citet{meurer_1999} after carrying out deep H$\alpha$ imaging of the disk.  

The \hi\ disk of NGC~2915 serves as an ideal tracer of the gravitational potential out to radii of $R\sim$~10~kpc, far beyond the visible radial extent of the stellar disk.  Due to the apparent lack of stars in its outer parts, the dynamics of the \hi\ disk are an almost direct tracer of the DM distribution.  \citet{meurer2} targeted NGC~2915 for DM studies and found the galaxy to be DM-dominated at nearly all radii with a total mass to $B$-band light ratio M$_{tot}$/L$_B$~$\gtrsim$~76~\msun/L$_{\odot}$ (assuming a distance of 5.1 Mpc).  They also found the galaxy to have a dense ($\rho_0~\approx~0.1$~\msun~pc$^{-3}$) and compact ($r_c\approx$~1~kpc) DM core.

%The observed spiral morphology as well as the kinematics of the NGC~2915's gas disk is difficult to understand.  The reason for the apparent absence of stars along the \hi\ spiral arms is not clear. Without an obvious energising source in the outer gas disk it is uncertain how the gas can sustain an average velocity dispersion $\sigma \gtrsim 10$ \kms.  
Several investigators have attempted to explain the observed \hi\ distribution of NGC~2915.  \citet{bureau_1999} studied the dynamics of the central \hi\ region and the spiral pattern of the outer disk.  For both they calculated a common, slow pattern speed of $\Omega_p=8.0\pm 2.4$~\kms\ which they associated with the figure rotation of a tri-axial DM halo.  They also proposed that some DM is distributed in the disk of NGC~2915, thereby making it gravitationally unstable to the formation of the observed spiral structure.  \citet{masset_bureau_2003}, using hydrodynamical simulations, further explored the ideas of \citet{bureau_1999}.  They showed that the observed spiral structure can be accounted for by either an unseen bar or a rotating tri-axial DM halo.  However, the mass of the required bar, $M_{bar}\sim 5\times 10^9$~\msun, is very large in comparison to the total stellar mass, thereby making the nature of such a bar problematic.  The required pattern speed of the tri-axial DM halo is significantly larger than those from numerical simulations.  \citet{masset_bureau_2003} disfavoured the external perturber scenario.   They also found that while a heavy disk is able to account for the main features of the observed \hi\ morphology, it fails to match the observed gas dynamics.  A satisfactory explanation for the various morphological and kinematic features of the \hi\ in NGC~2915 is therefore still lacking.  

In this paper we investigate the \hi\ dynamics and DM distribution of NGC~2915.  We use data from new \hi\ synthesis observations of NGC~2915 carried out using the Australian Telescope Compact Array as part of the Southern Hemisphere extension of The \hi\ Nearby Galaxy Survey.  Our \hi\ observations are significantly deeper and have better spatial resolution than any other \hi\ observations of NGC~2915.  These data are complemented by high-quality 3.6~\micron\ infrared observations of the stellar disk, carried out as part of the \emph{Spitzer} Infrared Nearby Galaxies Survey \citep[SINGS,][]{SINGS}.  The \hi\ observations are presented in Sec.~\ref{data_cube} while the \hi\ data products appear in Sec.~\ref{data_products}.  This paper focuses on the regular gas dynamics of NGC~2915.  In Sec.~\ref{vrot} a rotation curve out to $R\sim$~9.3~kpc is determined.  The far-infrared observations of the stellar disk are combined with our new \hi\ observations in Sec.~\ref{mass_modeling} to produce a mass model that determines the contributions at various radii of the stars, gas and DM.  We use two different parameterisations of the DM halo to reconstruct the observed rotation curve, one of which is the NFW halo \citep{NFW_1997} favoured by numerical simulations while the other is the observationally motivated pseudo-isothermal sphere.  Finally, in Sec.~\ref{conclusions}, we summarise our results and present our conclusions.
%-------------------------------------------------------------------------------------------------------------------------------------------------------------------
\section{\hi\ observations and data reduction}\label{data_cube}
\subsection{Data acquisition}
NGC 2915 was observed between 23~October~2006 and 2~June~2007 (project number C~1629) with six different ATCA configurations using all six antennas.  A single run consisted of a primary calibrator observation, regular secondary calibrator observations and source observations.  The calibrator sources PKS~1934--63 and PKS~0823--500 were used as primary and secondary calibrators respectively.  Each of the EW-352, 750D, 1.5B and 1.5C runs was approximately 12 hours long while 24 hours were spent in the 6A configuration.  Besides these data, archival data were also incorporated\footnote{Project number C191, principle investigator: Meurer.}.  \citet{meurer2} determined the H\,{\sc i} diameter of NGC 2915 to be $\sim$~0.32$^{\circ}$.   The H\,{\sc i} extent of NGC 2915 therefore falls well within the $\sim$~0.54$^{\circ}$  field-of-view of the ATCA dishes when observing at 21~cm.  For our observations, the telescope pointing centres were set to the optical centre of NGC 2915 ($\alpha_{2000}$~=~09$^{\mathrm{h}}$~26$^{\mathrm{m}}$~11.5$^{\mathrm{s}}$, $\delta_{2000}$~=~-76$^{\circ}$~37$'$~35$''$) with no mosaicking required.   The correlator was set to use 512 channels with a bandwidth of 8~MHz, centered at 1418~MHz.  The resulting velocity range is $-335$~\kms\ to 1353~\kms\ with an approximate channel spacing of 3.2~\kms.  Table \ref{ATCA_table} provides a summary of all our observing setups.

\begin{table}
\begin{center}
% use packages: array
\caption{Summary of NGC 2915 observing setups}
\label{ATCA_table}
\scriptsize{
\begin{tabular}{ccccccc}
\hline
\hline
\\
1	& 2 	& 3 	  & 4 	  & 5\\ 
Conf. & Date  &  Start & End & Dur. \\ 
 × & (yy-mm-dd) & (hh-mm-ss) & (hh-mm-ss) & (hrs) \\ 
\\
\hline
\\
 EW352 & 2006-10-23 & 11:28:55 & 23:17:45 & 10.39 \\
\\
  750D  & 2007-03-14 & 04:58:45 & 15:56:10 & 9.64 \\ 
\\
 1.5B  & 2006-11-24 & 19:29:05 & 06:51:35 & 9.95 \\  
\\
 1.5C   & 2007-05-02 & 04:37:25 & 15:36:05 & 9.64 \\ 
\\
6A   & 2007-02-10 & 10:34:35 & 21:32:25 & 9.64 \\
\\
 6A    & 2007-02-16 & 05:12:35 &  00:42:15& 17.06 \\
\\
\hline
 \end{tabular}}
\end{center}
\textbf{Comments on columns:} Column 1: ATCA configuration used; Column 2: date of observation (UT); Column 3/4: start/end of observations (UT); Column 5: time on source.
\end{table}

\subsection{H\,{\sc i} Data cubes}
The MIRIAD software package \citep{MIRIAD} was used to take the raw $uv$ data from the correlator through to the image analysis stage.    All velocities are measured relative to the barycentre rest frame.  The first five and last five correlator channels were flagged.  The data were then split into primary calibrator, secondary calibrator, and source subsets.  The required corrections to the antenna gains, delay terms and bandpass shapes were determined.  The time-varying phases and antenna gains were calibrated based on observations of the secondary calibrator.  %Corrections based on the primary calibrator gain corrections were applied to the secondary calibrator flux measurements.  Typical scaling factors were $\sim$ 1.2.  Finally, the corrections applied to the secondary calibrator were applied to the source data.

The calibrated source data were continuum-subtracted by fitting and subtracting a first-order polynomial to the line-free channels which, for our observations, were channels $50-100$ and $350-460$.  Having isolated the H\,{\sc i} signal, image cubes were produced from the continuum-subtracted $uv$ data.  Data cubes were produced using natural weighting as well as robust weighting with a Brigg's visibility weighting robustness parameter of 0.2.  Each of the dirty images was deconvolved using a Steer Clean algorithm \citep{steer_clean} to produce an output map of clean components.  Channels were either cleaned down to a flux cut-off of 2.5 times the typical r.m.s. of the flux in a line-free channel or for 50 000 iterations, whichever condition was met first.  The typical number of iterations was $\sim$ 20\,000 -- 30\,000.  After the deconvolution process, each of the clean components was convolved with a Gaussian approximation of the dirty beam.  The beam's full width at half maximum was 17$''\times  18.2''$ and $10.2''\times 10.2''$ for the naturally-weighted (NA) and robust-weighted (RW) data cubes respectively while the channel width was set to $dV=3.49$~\kms\ for both.  No Hanning smoothing was applied to either data cube.  For each cube, the noise in a line-free channel is Gaussian distributed with a mean of $\mu\sim-0.02$~mJy~beam$^{-1}$ and a standard deviation of $\sigma\sim$~0.60~mJy~beam$^{-1}$  for the NA cube, while $\mu\sim-0.004$~mJy~beam$^{-1}$ and $\sigma\sim 0.58$~mJy~beam$^{-1}$  for the RW cube.  

\subsection{H\,{\sc i} Moment maps}\label{moment_maps}
All H\,{\sc i} moment maps were generated using the Groningen Image Processing System (GIPSY, \citealt{gipsy}) together with the NA \hi\ data cube.  The entire \hi\ data cube was smoothed, using a Gaussian convolving function, to a resolution of 34$''\times 36.4''$.  A flux cut-off was applied to the smoothed cube at 2.5$\sigma$.  For each channel of the smoothed data cube, any flux that was clearly not spatially connected to, or associated with the galaxy emission was flagged and removed.  The blotted smoothed cube was then applied as a mask to the original $17''\times 18.2''$ cube.

The H\,{\sc i} total intensity and 2nd-order moment maps were constructed from the zeroth- and second-order moments of the data cube respectively.  A map in which each pixel is equal to $\sigma\sqrt{N}$, where $N$ is the number of un-blanked channels for a given line profile that would contribute to the moment maps, was created.  Such a construction was permitted since no Hanning smoothing was applied to the \hi\ data cube.  Dividing each pixel in the total intensity map by $\sigma\sqrt{N}$ produced a dimensionless signal-to-noise (S/N) map.  The average intensity of all the pixels in the total intensity map that had a corresponding signal-to-noise ratio in the range 2.75 $\leq$ S/N $\leq$ 3.25 was used as an intensity cut-off for the total intensity map.  This removed, in a robust manner, pixels from the outer disk that had low S/N values.  This new \hi\ total intensity map was then used as a mask on other H\,{\sc i} maps.

%Two different \hi velocity fields were constructed.  The first method involved calculating for each line profile its first moment or intensity-weighted-mean (IWM) velocity.  Such a method is simple to implement yet is significantly affected by profile asymmetries.  

The method of using Gauss-Hermite polynomials to parameterise line profiles of galaxies was introduced by \citet{GHpolynomials}.  While Gaussians are reasonable approximations to many realistic line profiles, higher order Gauss-Hermite polynomials are able to better capture profile asymmetries and non-Gaussian deviations.  The Hermite method has already been successfully implemented by various authors \citep{GHpolynomials,noordermeer2007,THINGS_deblok}.  We fitted a third-order Gauss-Hermite polynomial to each NA \hi\ data cube line profile to generate a so-called Gauss-Hermite \hi\ velocity field.  Three filters were used simultaneously when fitting the profiles: 1) profiles with fitted peak fluxes below 2.5$\sigma$ were excluded; 2) profiles with a fitted line width less than the channel width were excluded; and 3) fitted profile peaks had to be within the velocity range of the data cube.
%-----------------------------------------------------------------------------------------------------------------------------------------------------------------
\section{HI properties}\label{data_products}
\subsection{Channel maps}\label{channel_maps}
A grey-scale representation of the $17''\times18.2''$ resolution channel maps is shown in Fig. \ref{channels}.  $\sigma\sim 0.6$~mJy in all channels.  \hi\ structures are visible at large and small scales.  At large scales, the emission clearly exhibits the usual pattern of a rotating disk.  The \hi\ disk of NGC 2915 exhibits clear spiral structure (c.f. \hi\ total intensity map, Fig. \ref{mom0}).  On small scales, the gas in the central regions forms two dominant over-densities in nearly all of the channels, with fainter emission in-between them.

\begin{figure*}
	\centering
	\includegraphics[width=1.8\columnwidth, angle=0]{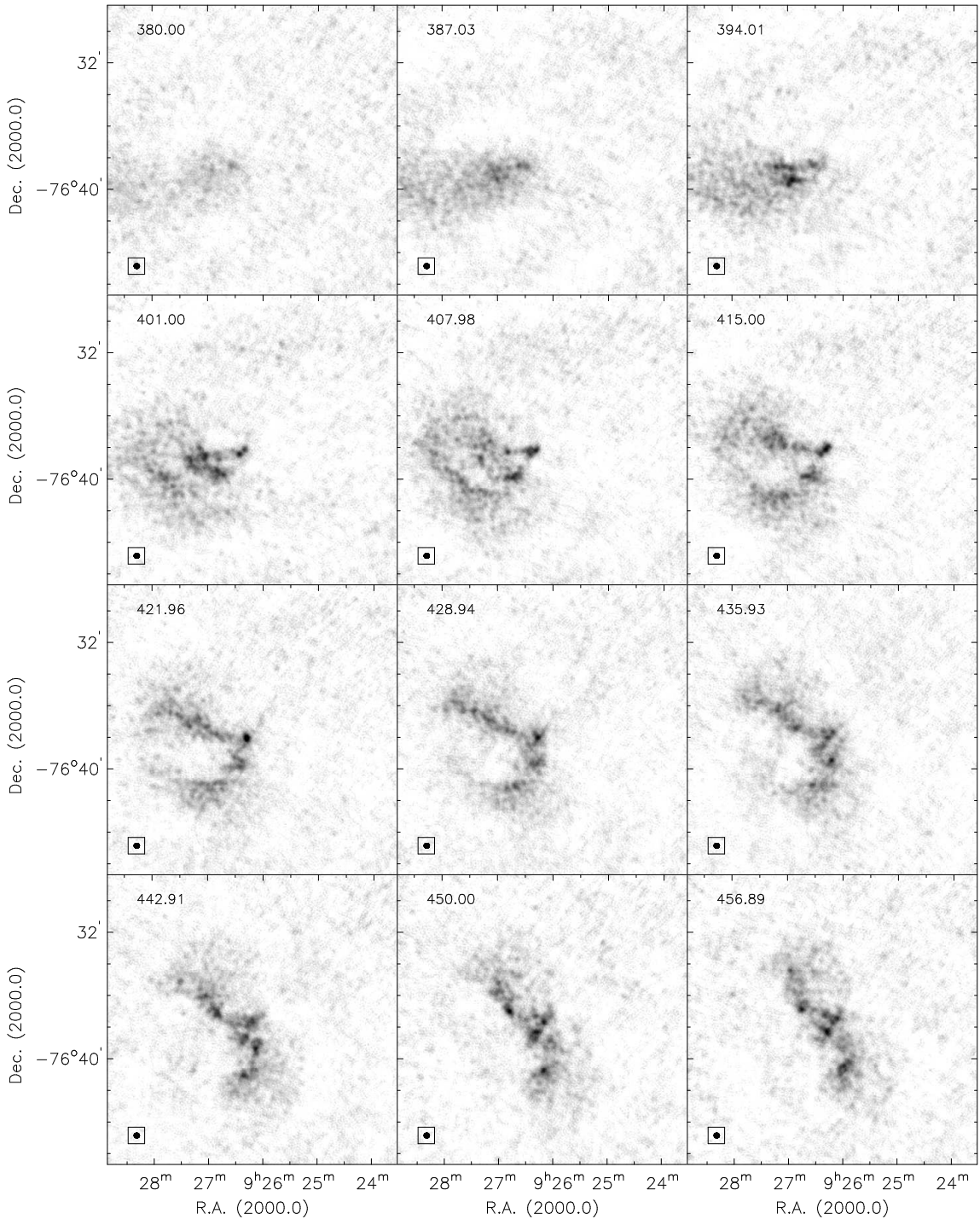}
	\caption{Channel maps of NA \hi\ data cube.  Every second channel is shown.  The heliocentric radial velocity (km s$^{-1}$) of each channel is shown in the upper left corner.  The half-power-beam-width is shown in the bottom left corner.  Grey-scale range is from -0.6 mJy beam$^{-1}$ to 10.0 mJy beam$^{-1}$.  The r.m.s. noise in a channel is $\sim$ 0.6 mJy beam$^{-1}$.}
	\label{channels}
\end{figure*}
\addtocounter{figure}{-1}

\begin{figure*}
	\centering
 	\includegraphics[width=1.8\columnwidth, angle=0]{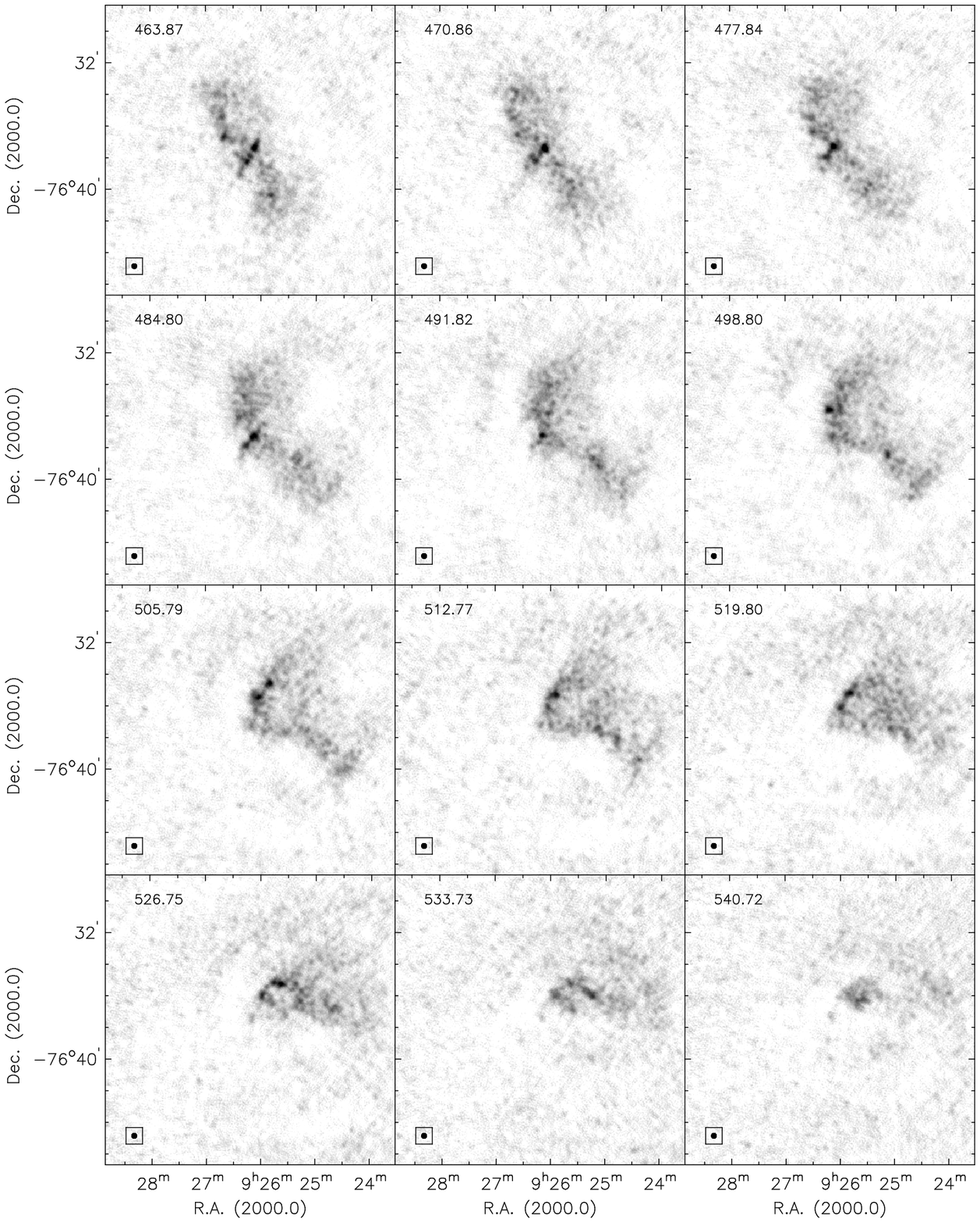}
	\caption{Continued.}
\end{figure*}
%-----------------------------------------------------------------------------------------------------------------------------------------------------------------
\subsection{Global profile}\label{global_profile}
\begin{figure}
	\begin{centering}
	\includegraphics[angle=0,width=1.0\columnwidth]{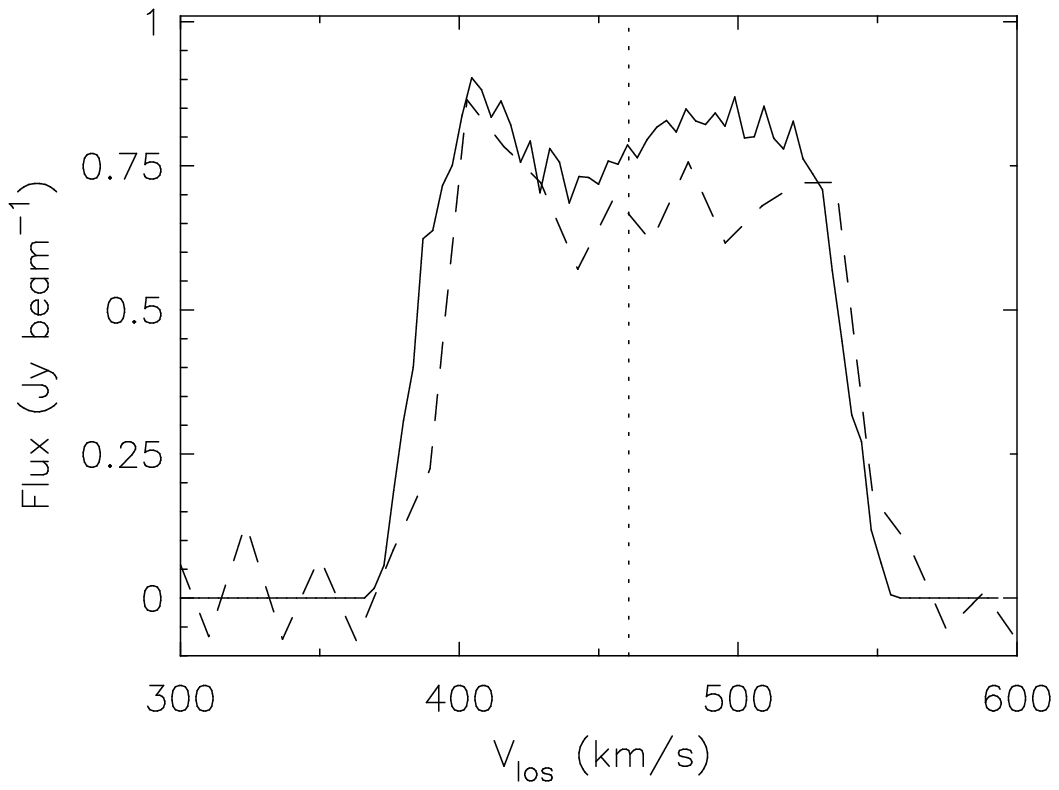}
	\caption{Global H\,{\sc i} profile from our observations (solid curve) and HIPASS (dashed curve).  The dotted vertical line shows the systemic velocity as calculated using Eqn \ref{vsysest}.}
	\label{global_HI_profiles}
	\end{centering}
\end{figure}

A global \hi\ profile, extracted from the blotted NA \hi\ data cube by summing the \hi\ emission in each velocity channel, is presented in Fig.~\ref{global_HI_profiles}.  Using a synthesis array to estimate the total \hi\ flux of a galaxy can result in an underestimate due to missing short baselines.  The shortest baseline used for our observations was 31~m (as part of the EW352 antenna configuration), implying that the observations are less sensitive to structures larger than $\sim$ 23$'$.  \citet{meurer2}, however, estimated the H\,{\sc i} extent of NGC~2915 to be $\sim$ 19$'$.  The amount of H\,{\sc i} missed in the global H\,{\sc i} profile is therefore expected to be small.  In Fig.~\ref{global_HI_profiles} we compare our determination of the global \hi\ profile for NGC~2915 to the HIPASS global \hi\ profile from \citet{HIPASS_1000}.  The HIPASS data are calibrated with 8$'$ spatial binning and 13 km s$^{-1}$ channel width.  Due to the large HIPASS beam, as well as the fact that the Parkes telescope is a single dish telescope, the amount of flux missed in the HIPASS observations should be negligible.  It is clear from Fig.~\ref{global_HI_profiles} that no significant amount of flux has been missed by our observations.

The resulting global \hi\ profile is double-horned and asymmetric.  From this profile we derived the profile widths, the systemic velocity, and the total H\,{\sc i} mass.  The profile widths were calculated as the differences between the high and low velocities of the galaxy at 20$\%$ and 50$\%$ of the peak flux density (V$_{high}^{20\%}$, V$_{low}^{20\%}$, V$_{high}^{50\%}$, V$_{low}^{50\%}$ respectively).  The profile widths were not corrected for inclination.  Two systemic velocity estimates were determined: 1) from the global profile, at the midpoint of the two velocities at the 50$\%$ level, 2)  from the equation
%The uncertainties in the W$_{50}$ and W$_{20}$ parameters were estimated by allowing the the 20$\%$ and 50$\%$ flux density levels to vary within the typical channel RMS noise and then taking the difference between the largest and smallest profile widths.
\begin{equation}
V_{sys}=0.25\times(V_{high}^{20\%}+ V_{low}^{20\%}+ V_{high}^{50\%}+ V_{low}^{50\%}). 
\label{vsysest} 
\end{equation}
Finally, the total H\,{\sc i} mass was calculated as 
\begin{equation}
M_{HI}=2.36\times 10^5\times D^2\times\int F dV,
\end{equation}
where $D$ is the distance to the galaxy in Mpc and $\int F dV$ is the total H\,{\sc i} line flux in Jy~km~s$^{-1}$.  This determination of the total H\,{\sc i} mass assumes that the H\,{\sc i} is optically thin.  Throughout this paper we adopt the $D\sim 3.78$ Mpc distance determination of \citet{karachentsev_catalog}.  Table~\ref{HI_global_profile_table} lists our determinations of the above mentioned quantities.   Assuming a distance of $D=5.1$~Mpc, \citet{meurer2} measured a total \hi\ mass of $M_{HI}\sim1.27\times 10^9$~\msun\ for NGC~2915.  Using a distance of $D=3.78$~Mpc, their total mass estimate reduces to $M_{HI}\sim 7 \times10^8$~\msun, consistent with our estimate of $M_{HI}\sim 4.4\times 10^8$~\msun.

\begin{table}
\begin{center}
\caption{Quantities derived from the global H\,{\sc i} profile.}
\scriptsize{
\begin{tabular}{ccccc}
\hline
\hline
\\
1 & 2 & 3 & 4 & 5 \\ 
$M_{HI}$ & $W_{50}$  & $W_{20}$ & $V_{sys}^{50}$ & $V_{sys}^{\mathrm{Eqn. 1}}$ \\ 
(10$^8$ M$_{\odot}$) & (km s$^{-1}$) & (km s$^{-1}$) & (km s$^{-1}$) & (km s$^{-1}$) \\
\\
\hline
\\
4.4  & 186.8 & 170.0 & 460.5 & 460.8\\
\\
\hline
\end{tabular}}
\label{HI_global_profile_table}
\\
\end{center}
\textbf{Comments on columns:} Column 1: total H\,{\sc i} mass; Column 2: velocity width at 50$\%$ of the peak flux; Column 3: velocity width at 20$\%$ of the peak flux; Column 4: systemic velocity from W$_{50}$ midpoint; Column 5: systemic velocity from Eqn. \ref{vsysest}.
\end{table}

%This plot was made with /Volumes/PORTABLE/science/ngc2915/profiles/global_HI/global_profile.py
%-----------------------------------------------------------------------------------------------------------------------------------------------------------------
\subsection{Total intensity map}\label{total_intensity}
\begin{figure}
	\centering
	\includegraphics[angle=0,width=1.0\columnwidth]{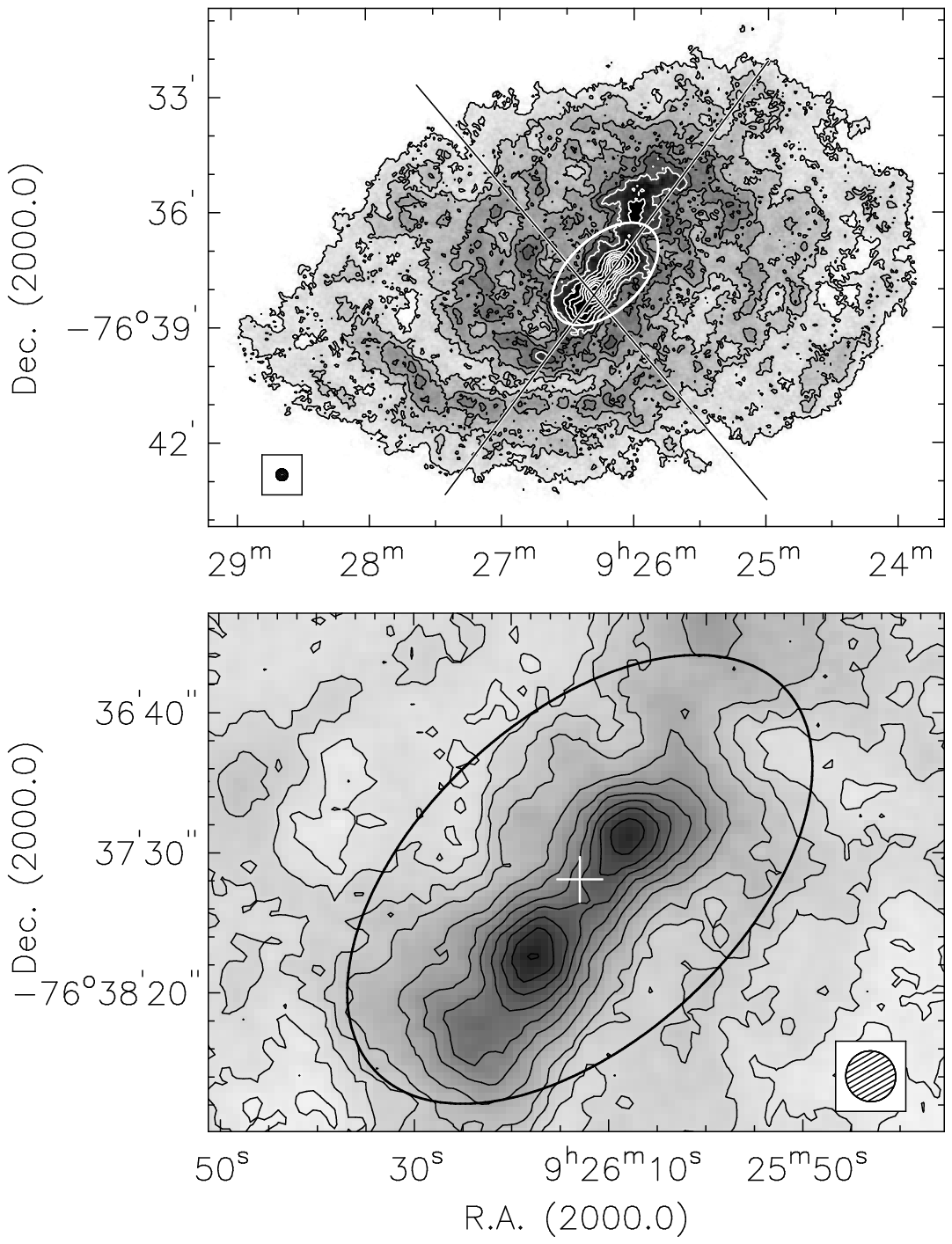}
	\caption{Grey-scale plot of the \hi\ total-intensity map (top panel) with a zoom-in of the central region (bottom panel).  In both panels the intensity scale runs from 5~mJy~beam$^{-1}$~(white) to 0.2~Jy~beam$^{-1}$ (black).  Contour levels run from  10~--~60~mJy~beam$^{-1}$ in steps of 10~mJy~beam$^{-1}$ and from 80~--~230~mJy~beam$^{-1}$ in steps of 20~mJy~beam$^{-1}$.  The position of the photometric centre estimated using the 3.6~\micron\ image is marked with a cross.  The hatched circles in the lower corners represent the half power beam width of the synthesised beam.  The black and white ellipses have a semi-major axis length $a=50''$, an inclination $i=55^{\circ}$, a position angle $PA=317^{\circ}$ and a centre position equal to our determination of the photometric centre.  The solid black lines represent the position-velocity slices shown in Fig.~\ref{pv_slices}.}
	\label{mom0}
\end{figure}

\begin{figure}
	\centering
	\includegraphics[angle=0,width=1.0\columnwidth]{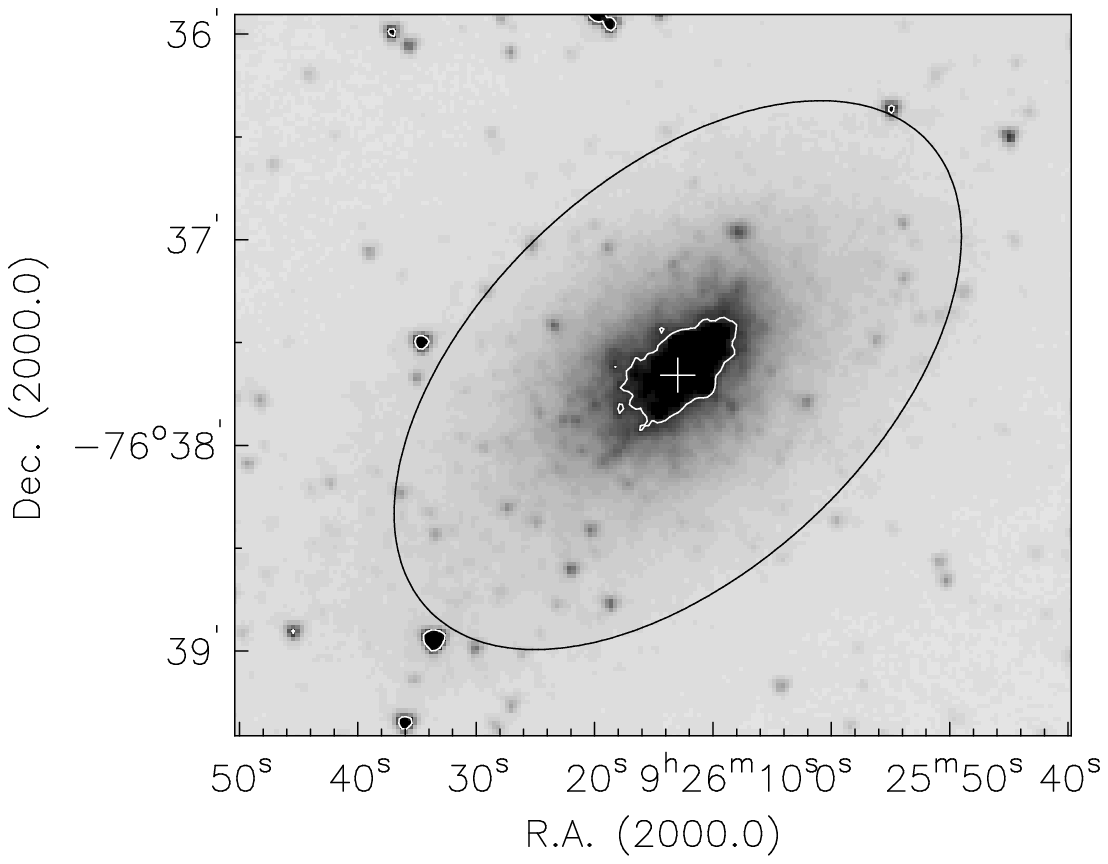}	
	\caption{Grey scale plot of the 3.6~\micron\ IRAC Spitzer image.  The intensity scale runs from -0.2~MJy~ster$^{-1}$~(white) to 2.0~MJy~ster$^{-1}$~(black).  The single white contour is at a level of 1.2~MJy~ster$^{-1}$ and represents the average flux of one of three flux annuli used to estimate the position of the photometric centre.  The estimated photometric centre is marked with a cross.  The black ellipse is the same as that shown in Fig.~\ref{mom0}}
	\label{IRAC_image}
\end{figure}

The full \hi\ total intensity map extracted from the NA \hi\ data cube is shown in the upper panel of Fig. \ref{mom0}.  The edge of the \hi\ disk (the outer-most contour level) lies at $R\sim$~510$''$ and therefore extends out to $R\sim$~20 $B$-band scale lengths.  For comparative purposes the IRAC 3.6~\micron\ image of the stellar disk is shown in Fig. \ref{IRAC_image}.  To determine the photometric centre we fitted ellipses to three 3.6~\micron\ flux density annuli near the edge of the old stellar disk.  A 3.6~\micron\ flux density contour at a level of 1.2~MJy~ster$^{-1}$, the average flux density of the second annulus, is shown in Fig.~\ref{IRAC_image}.  The average flux densities of the other two annuli are 0.8 and 1.7~MJy~ster$^{-1}$.  The average centre position of the ellipses fitted to the flux density annuli, $\alpha_{2000}$~=~09$^\mathrm{h}$~26$^\mathrm{m}$~12.611$^\mathrm{s}$, $\delta_{2000}$~=~$-76$\deg~37$'$~37.80$''$,  was used to estimate the position of the photometric centre (shown as a cross in Fig.~\ref{IRAC_image}).  For comparison, the 2MASS determination of the centre of the system is  $\alpha_{2000}$~=~09$^\mathrm{h}$~26$^\mathrm{m}$~11.53$^\mathrm{s}$, $\delta_{2000}$~=~$-76$\deg~37$'$~34.80$''$ \citep{skrutskie_2006}.  To facilitate length-scale comparisons between IR and \hi\ images, an ellipse with semi-major axis length $a=50''$, roughly equal to the $R$-band $R_{25}$ radius determined by \citet{meurer1}, is included in the figures.  In their study of the \hi\ content of 73 late-type dwarf galaxies, \citet{WHISP_swaters} found that the ratio of the \hi\ extent to the optical diameter, defined as 6.4 disk scale lengths, is 1.8 $\pm$ 0.8 on average.  The same ratio for NGC~2915 is $\sim 3.4$, close to double that value.

\begin{figure}
	\centering
	\includegraphics[angle=0,width=1.0\columnwidth]{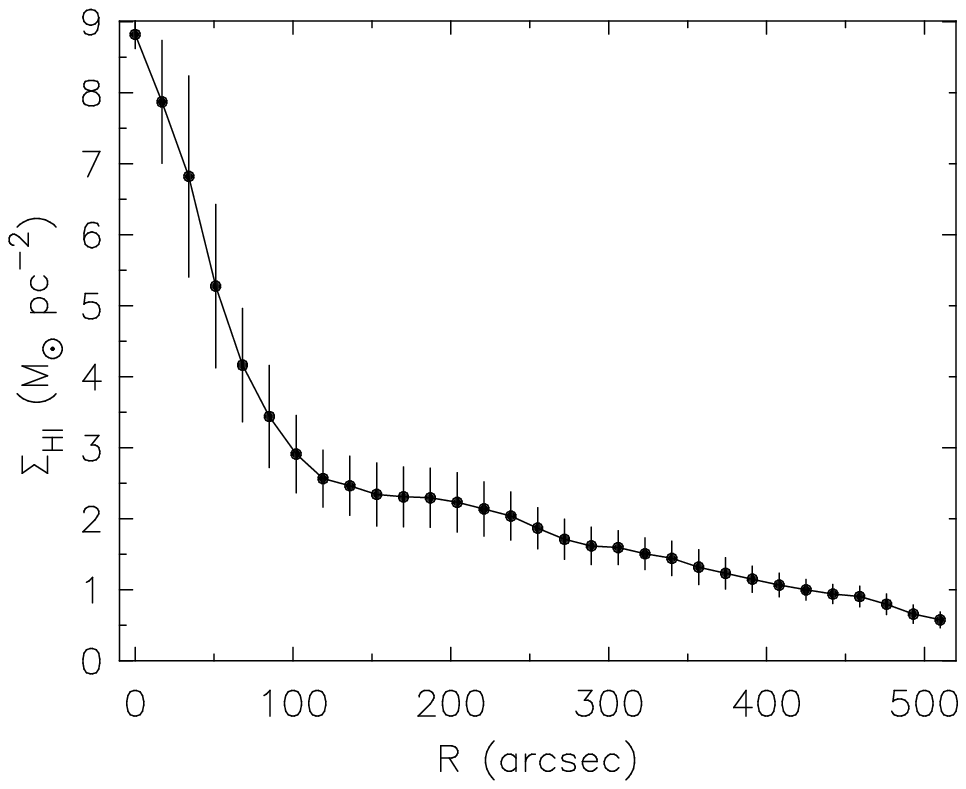}	
	\caption{\hi\ surface density radial profile.  All surface densities have been inclination-corrected.  Error bars represent the r.m.s. spread of the surface densities in each azimuthally averaged ring.}
	\label{surf_dens_prof}
\end{figure}

Our high-resolution \hi\ observations resolve the central regions of NGC~2915 into two distinct \hi\ concentrations (Fig.~\ref{mom0}, lower panel).  Their centres are separated by $\sim$~60$''$ (1.1~kpc) and have a combined mass of $\sim 1.9\times 10^7$~\msun.  Also visible is a plume-like \hi\ feature to the North-West of these central \hi\ concentrations with a mass of $\sim 2.8\times 10^7$~\msun.  The observed spiral structure of the outer disk is asymmetric: the arm beginning at the North-Western central \hi\ concentration can be traced over $\sim$ 200$^{\circ}$ in azimuthal angle while the arm beginning at the South-Eastern \hi\ concentration can be traced for only $\sim$ 120$^{\circ}$.  From the H\,{\sc i} total intensity map an azimuthally averaged, inclination-corrected H\,{\sc i} surface density profile was constructed.  Each ring has a width of $dR=17''$, a position angle (measured anti-clockwise from North to the receding major axis) of $PA=285^{\circ}$ and an inclination of $i=55$\deg.  The profile is shown in Fig. \ref{surf_dens_prof}.

In an attempt to better constrain the \hi\ distribution of NGC~2915, an \hi\ total intensity map was extracted from the RW \hi\ data cube.  For the sake of brevity this map is not shown here, yet it suffices to say that the higher spatial resolution confirms the above-mentioned nature of the central \hi\ concentrations.  However, the very low S/N ratio in the outer parts of this map yields it unsuitable for the study of low column density \hi\ features, with only the central-most \hi\ features standing out about the noise.  This map does not allow for a more accurate study of the \hi\ distribution and is therefore not discussed further.   %\citet{WHISP_swaters} found the average \hi\ mass of their sample 73 late-type galaxies to be $\sim 13.0\times10^8$ \msun.  Thirty one out of their 73 galaxies did, however, have a total \hi\ mass $\le$ 5.0 $\times$ 10$^8$ \msun.

%from /Volumes/PORTABLE/science/ngc2915/ellint2
%------------------------------------------------------------------------------------------------------------------------------------------------------------------------------
\subsection{Velocity field}\label{vel_field}
\begin{figure}
	\centering
	\includegraphics[angle=0,width=1.0\columnwidth]{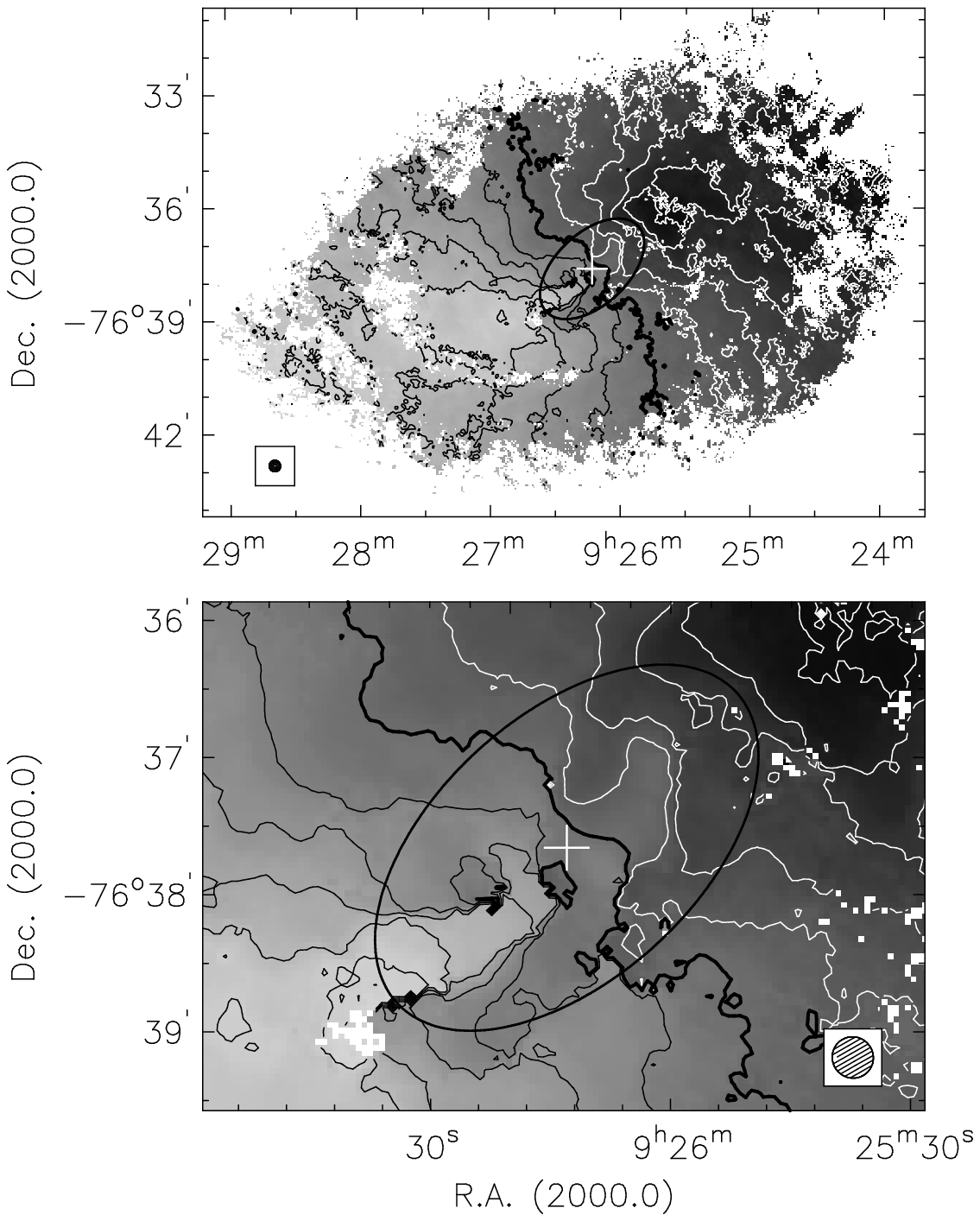}
	\caption{Top panel: Grey-scale plot of the third-order Gauss-Hermite H\,{\sc i} velocity field.  Bottom panel: A zoom-in of the central region.  The intensity scale runs from 300 -- 600 \kms.  Contours are separated by 15 km s$^{-1}$ with the thick contour marking the systemic velocity at 465 km s$^{-1}$ derived by fitting a tilted ring model to the velocity field.  White and black contours represent receding and approaching halves of the galaxy respectively.  The hatched circles in the lower corners represent the half power beam width of the synthesised beam.  The ellipse is the same as that shown in Fig. \ref{IRAC_image}.}
	\label{mom1}
\end{figure}

The third-order Gauss-Hermite velocity field extracted from the NA \hi\ data cube is shown in Fig.~\ref{mom1}.  The blank pixels in this map are those that were rejected by the fitting routine according to the criteria mentioned in Sec. \ref{moment_maps}.  \citet{THINGS_deblok} showed that the Gauss-Hermite parameterisation of a line profile gives a robust estimate of the peak-associated velocity of the profile.  Inner contours running approximately parallel to the kinematic minor axis are caused by $V(R)\propto R$ rotation while the outer contours running radially away from the centre are due to a constant rotation speed.  The overall ``S-shaped'' distortion of the contours is indicative of a kinematic warp within the disk.  Small wiggles along the outer contours are caused by streaming gas motions.  The sharp kinks at inner radii suggest the presence of non-circular velocity components within the gas.  Finally, differences in the shapes of the iso-velocity contours on the receding and approaching halves of the galaxy suggest a certain degree of kinematic lopsidedness within the galaxy.  

A third-order Gauss-Hermite velocity field was also extracted from the RW \hi\ data cube.  Due to the high noise levels in the RW \hi\ cube, most line profiles were rejected by the fitting filters and hence not fitted.  The general characteristics of the fitted peaks of the remaining profiles are very similar to those of the NA velocity field, differing on average by $\sim 5$~\kms.   Rather than showing this \hi\ velocity field here, a tilted ring model is fitted to it in Sec.~\ref{TR_output} to demonstrate that the resulting rotation curve is very similar to the one derived using the NA third-order Gauss-Hermite velocity field.   Motivated by this result is the decision to use only the NA velocity field for our kinematic analyses.

%------------------------------------------------------------------------------------------------------------------------------------------------------------------------------
\subsection{Second-order moment map}\label{disp_map}
\begin{figure}
	\centering
	\includegraphics[angle=0,width=1.0\columnwidth]{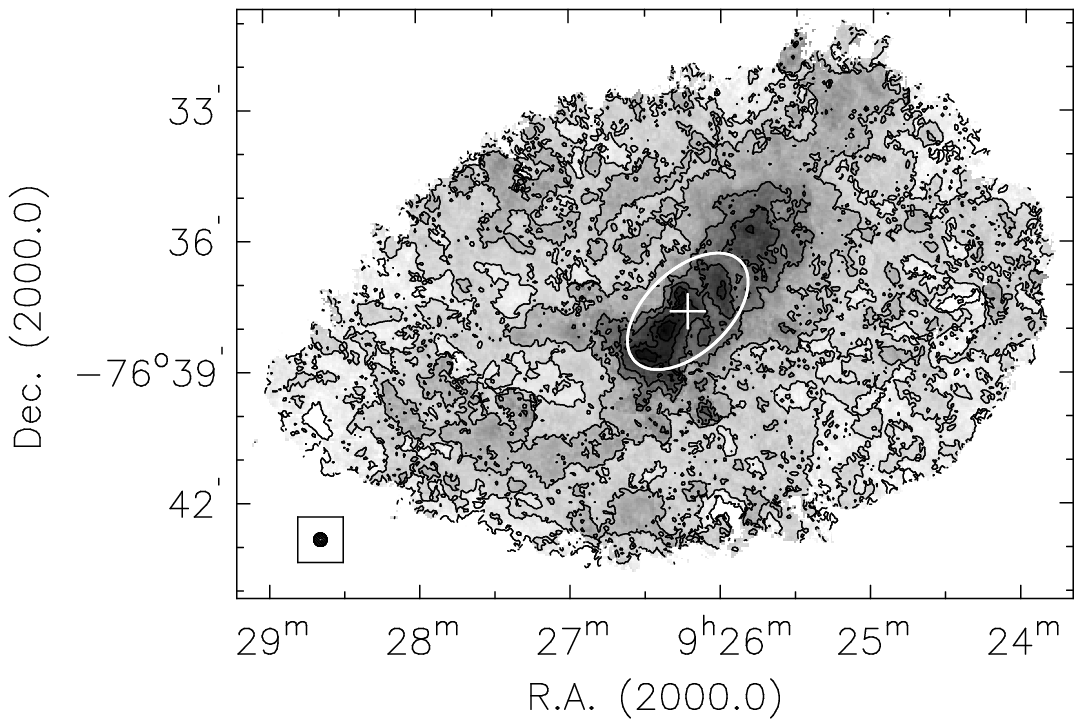}
	\caption{Grey-scale plot of the second-order H\,{\sc i} moment map.  The grey scale runs from 6~\kms~(white) to 30~\kms~(black).  Contour levels are at 6, 9, 12, 18, 22, 26, 28, 30~\kms. The hatched circle in the lower left corner represents the half power beam width of the synthesised beam.  The ellipse is the same as that shown in Fig. \ref{IRAC_image}.}
	\label{mom2}
\end{figure}

\begin{figure}
	\begin{centering}
 	\includegraphics[width=1.0\columnwidth]{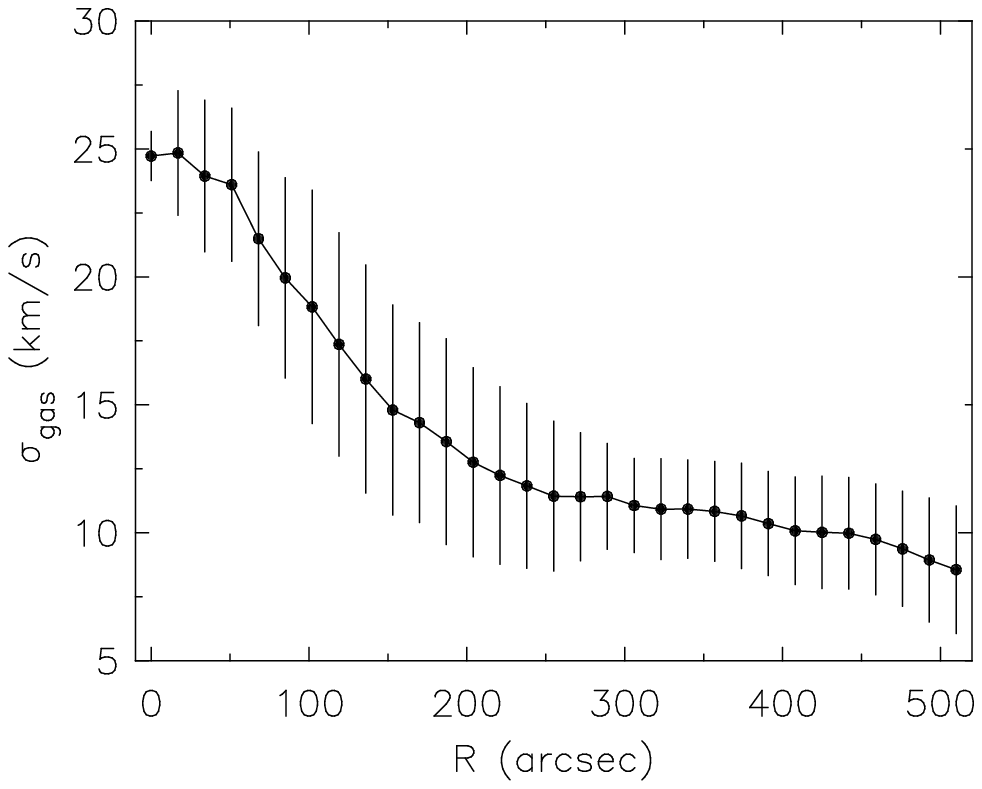}
	\caption{Second-order H\,{\sc i} moment map radial profile.  Error bars represent the r.m.s. spread of the 2nd-order moments in each azimuthally averaged ring.}
	\label{disp_prof}
	\end{centering}
\end{figure}

The second-order \hi\ moment map extracted from the NA \hi\ data cube is shown in Fig.~\ref{mom2}.  The highest second-order moments are found near the stellar core of the galaxy.  The radial profile of the second-order moment map is shown in Fig.~\ref{disp_prof}.  If the second-order moments are interpreted as a measure of the gas velocity dispersion, then the average outer disk dispersion of $\sigma_{gas}\sim$ 10 km s$^{-1}$ is close to the constant gas velocity dispersion value of  $\sigma_{gas}=11$~\kms\ found by \citet{leroy_THINGS} for a sample of THINGS dwarf galaxies.

%------------------------------------------------------------------------------------------------------------------------------------------------------------------------------
\subsection{Non-Gaussian line profiles}\label{line_profiles}
\begin{figure}
	\begin{centering}
	\includegraphics[angle=0,width=1.0\columnwidth]{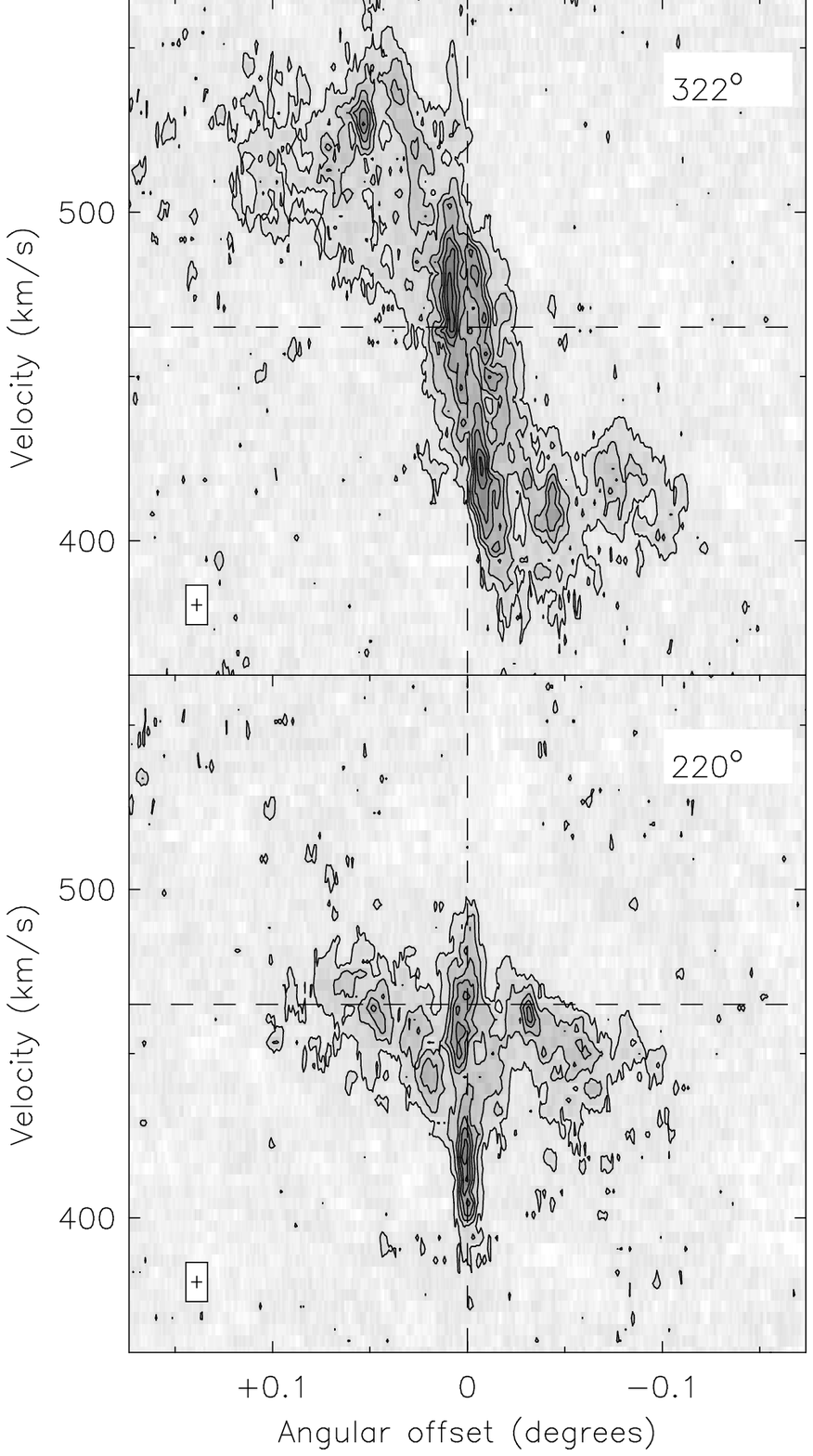}
	\caption{Position-velocity diagrams extracted along 1) a position angle of 322$^{\circ}$ (top) and 2) a position angle of 220$^{\circ}$ (bottom).  The slice positions are shown in the \hi\ total intensity map (Fig.~\ref{mom0}).  The thickness of the slices is that of a single pixel in the data cube (2.5$''$) and the velocity resolution is 3.5 km s$^{-1}$.  The intensity scale runs from --3$\sigma$ to (white) 40$\sigma$ (black).  Contours start at 2$\sigma$ in steps of 3$\sigma$.  The dashed horizontal line marks the systemic velocity at $V_{sys}=465$~\kms.}
	\label{pv_slices}
	\end{centering}
\end{figure}

Position-velocity slices taken though the centre of the NA \hi\ data cube betray the presence of broad and multi-component line profiles at inner and outer radii.  Figure \ref{pv_slices} shows two position-velocity slices, the positions of which are shown in the \hi\ total intensity map (Fig. \ref{mom0}, top panel).  The first slice was taken at a position angle of 322\deg, through both of the central \hi\ concentrations.  This slice suggests that the galaxy contains a fast-rotating gas component near its centre, clearly manifesting itself as a sharp velocity spike with a peak velocity of $V\sim$ 550 \kms.  Line profiles near the centre of this position-velocity slice show significant departures from Gaussianity.

This position-velocity slice also reveals asymmetric line profiles on the receding side of the galaxy, seen in the upper-left quadrant ($V>V_{sys}=465$~\kms, angular offset $<0$) as an ``\hi\ beard'' that is lagging in velocity relative to the main disk.  The beard emission spatially coincides with the plume-like \hi\ feature seen in the \hi\ total intensity map.  The estimated mass of this morphological feature, $\sim 2.8\times 10^7$~\msun, is 5.6$\%$ of total \hi\ mass.  One interpretation of this kinematically anomalous gas component is that it is associated with pristine gas being accreted from the nearby inter-galactic space onto the outer disk of NGC 2915.  If confirmed, this would be the first evidence of such an event in a dwarf system like NGC~2915. 

The second position-velocity slice was extracted almost parallel to the minor axis, through the South Eastern central \hi\ concentration which contains many double--component profiles.  Line profiles in this central region are split by $\sim~30$~\kms\ on average, implying that the sharp rise seen toward the centre of the second-order \hi\ moment map (Fig.~\ref{mom2}) over-estimates the true \hi\ velocity dispersions of the inner disk.  Since the disturbed \hi\ line profiles are located within $R\sim 150''$ of the centre of NGC 2915, it is plausible that the central gas dynamics of NGC~2915 are largely dictated by the stellar winds of its young stellar population.  In a forthcoming paper (Elson et al. 2010, in prep.) we present detailed modeling results of the central \hi\ dynamics.

%-----------------------------------------------------------------------------------------------------------------------------------------------------------------
\section{\hi\ dynamics}\label{vrot}
The H\,{\sc i} line data allow us to study the H\,{\sc i} gas dynamics.  In this section we derive a rotation curve for NGC~2915.

\subsection{Tilted ring model}
The standard method of producing a rotation curve involves fitting a tilted ring model to a velocity field \citep{begemann_phd_thesis}.  NGC~2915's intermediate inclination of $i\sim$55$^{\circ}$ makes it a well-suited candidate for such an analysis.  The method involves modeling the galaxy as a set of concentric rings within which the gas orbits about the kinematic centre.  Each ring has a set of defining parameters: central coordinates $X_c$ and $Y_c$, inclination $i$, systemic velocity $V_{sys}$, position angle $PA$ (in the sky plane\footnote{Measured anti-clockwise from North to the receding semi-major axis.}), and the rotation velocity $V_{rot}$ of the material within the ring.  In the case that expansion velocities within the disk are ignored, the standard algorithm carries out a least squares fit to
\begin{equation}
V_{los}(x,y)=V_{sys}+V_{rot}\sin(i)\cos(\theta),
\label{vlos}
\end{equation}
where $V_{los}$ is the line-of-sight velocity, $x$ and $y$ are rectangular coordinates on the sky and $\theta$ is the angle from the major axis in the galaxy plane.  $\theta$ is related to the position angle $PA$ in the sky plane by
\begin{eqnarray}
\cos(\theta)&=&{-(x-X_c)\sin(PA)+(y-Y_c)\cos(PA)\over r},\\
\sin(\theta)&=&{-(x-X_c)\cos(PA)+(y-Y_c)\sin(PA)\over r\cos(i)},
\label{vlos}
\end{eqnarray}
where $r$ is the radius of the ring in the galaxy plane.

The GIPSY task ROTCUR  \citep{begemann_phd_thesis} was used to fit tilted ring models to the third-order Gauss-Hermite H\,{\sc i} velocity field extracted from the NA \hi\ data cube.  Rings width of $dr=17''$ were used to ensure that adjacent rings were largely independent of one another.  When fitting Eqn.~\ref{vlos} to the data, each datum was weighted by $|\cos(\theta)|$ so that points closer to the major axes held more weight.  All points within 10$^{\circ}$ of either side of the minor axes were excluded from the fit.  Both sides of the galaxy were used for the fitting. 

\subsection{Fitting procedure}
\begin{figure*}
	\begin{centering}
 	\includegraphics[width=2.0\columnwidth]{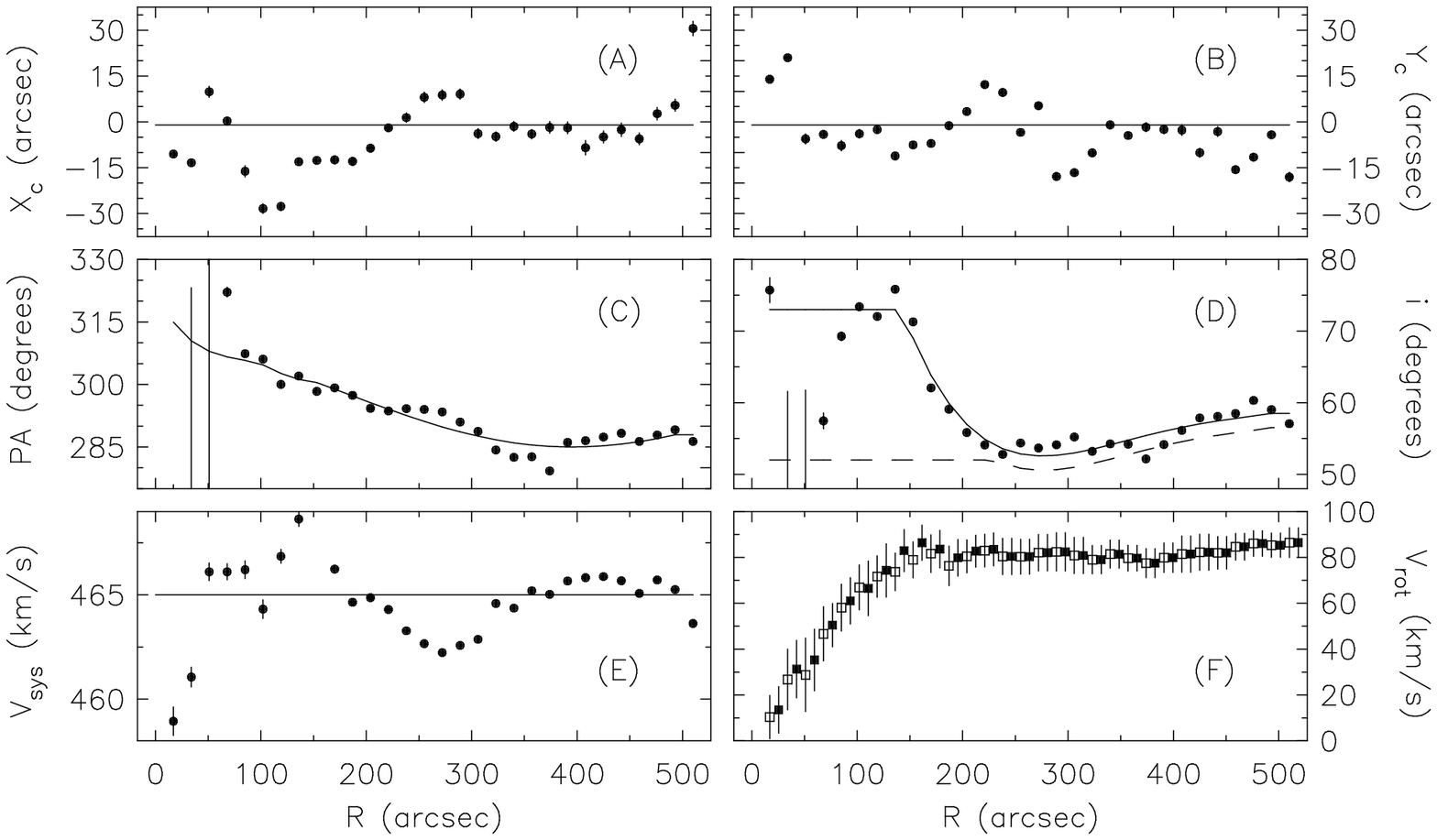}
	\caption{Radial variations of tilted ring parameters fitted to the third-order Gauss-Hermite velocity field.  Panels are A: central X positions, B: central Y positions, C: position angles, D: inclinations, E: systemic velocities, and F: circular rotation velocities.  Filled circles show the radial variations just before the parameter profiles were smoothed and fixed in order to derive the final rotation curves.  The smoothed profiles used to construct models SI and CI are shown as solid curves except for the model CI inclination profile which appears as a dashed curve (offset by -2$^{\circ}$ for the sake of clarity).  Panel F shows the final rotation curves derived for models SI and CI (open and filled squares respectively).  For the sake of clarity the rotation curves have been offset by 8.5$''$.  Error bars represent the r.m.s. spread of the velocities in a given ring.}
	\label{TR_results}
	\end{centering}
\end{figure*}

Each tilted ring should be centred on the dynamical centre of the galaxy, assuming it is constant, which is usually estimated by the least-squares fitting algorithm together with the other tilted ring parameters.  In the case of NGC~2915, however, the kinematic centre determinations from ROTCUR varied with radius in a non-systematic manner.  We therefore used the photometric centre of the 3.6~\micron\ emission as the kinematic centre.  Using the photometric and kinematic centres interchangeably in this manner is reasonable if one assumes that the stars lie at the bottom of the gravitational potential.   We carried out a single ROTCUR iteration with all parameters allowed to vary freely.  The resulting $X_c$ and $Y_c$ values are shown as filled black circles in panels A and B respectively of Fig. \ref{TR_results}.  These fitted $X_c$ and $Y_c$ positions deviate on average by 8.3$''$~$\pm$~$0.3''$ and 7.2 $''$~$\pm$ 0.25$''$ (152~pc and $\sim$ 132~pc) respectively from the photometric centre, thereby placing them well within a single half power beam width.  This average deviation is consistent with the results of \citet{trachternach_THINGS} who found that approximately 50$\%$ of their kinematic centres derived for $\sim$~1000 individual tilted rings fitted to the THINGS \hi\ velocity fields differed by less than a beam width from their best centre estimates.  In the case of the THINGS sample, one beam width corresponds to $\sim$~200~pc on average.  Throughout the tilted ring fitting procedure, $X_c$ and $Y_c$ were therefore each fixed to the position of the photometric centre (solid lines in panels A and B of Fig. \ref{TR_results})

The remaining tilted ring parameters ($V_{sys}$, $PA$, $i$, $V_{rot}$) were derived iteratively.  We first allowed all parameters to vary freely.  This gave a general feel for the parameter behaviour with $r$.  Little scatter was seen in the radial run of $V_{sys}$, which was then fixed to the average value of 465~\kms\ in subsequent iterations (compare with the estimate of $V_{sys}=460.8$~\kms\ derived from the global \hi\ profile using Eqn.~\ref{vsysest}).  The $PA$, $i$, and V$_{rot}$ parameters were allowed to vary freely in the subsequent iterations.  The $PA$ and $i$ runs showed relatively little scatter except for a clear change from outer to inner disk beginning at $R$ $\sim$ 240$''$ where the inclination value changed from $i\sim$ ~55$^{\circ}$ to $i\sim$~75$^{\circ}$.  Our kinematically derived position angles were consistent with the position angles of ellipses fitted to flux contours in the H\,{\sc i} total intensity map.  Several more iterations with various combinations of fixed and free parameters were carried out in order to check the stability of the radial parameter profiles.  The sharp rise in $i$ for R $\lesssim$ 240$''$ always remained with little scatter.  

\begin{figure}
	\begin{centering}
 	\includegraphics[width=1\columnwidth]{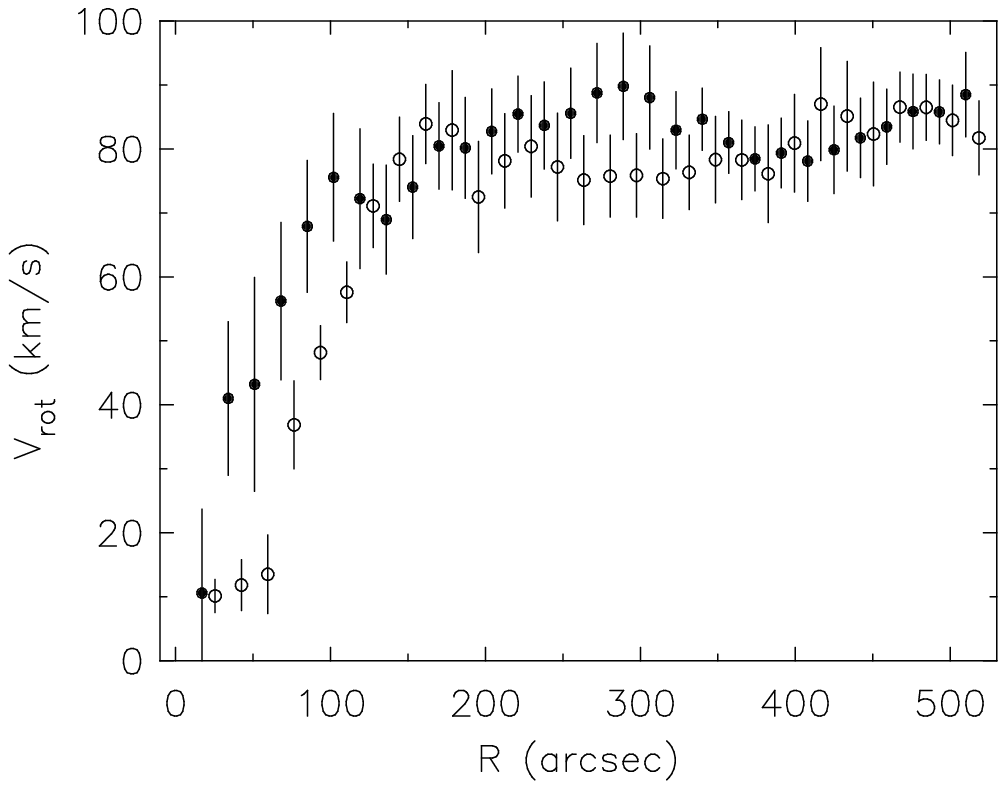}
	\caption{Rotation curves from tilted ring model SI derived separately for the approaching (filled circles) and receding (open circles) sides.  For the sake of clarity the rotation curve of the approaching side has been offset by 8.5$''$.  Error bars represent the r.m.s. spread of the velocities in a given ring.}
	\label{2vrots}
	\end{centering}
\end{figure}

\begin{figure*}
	\begin{centering}
 	\includegraphics[width=1.85\columnwidth]{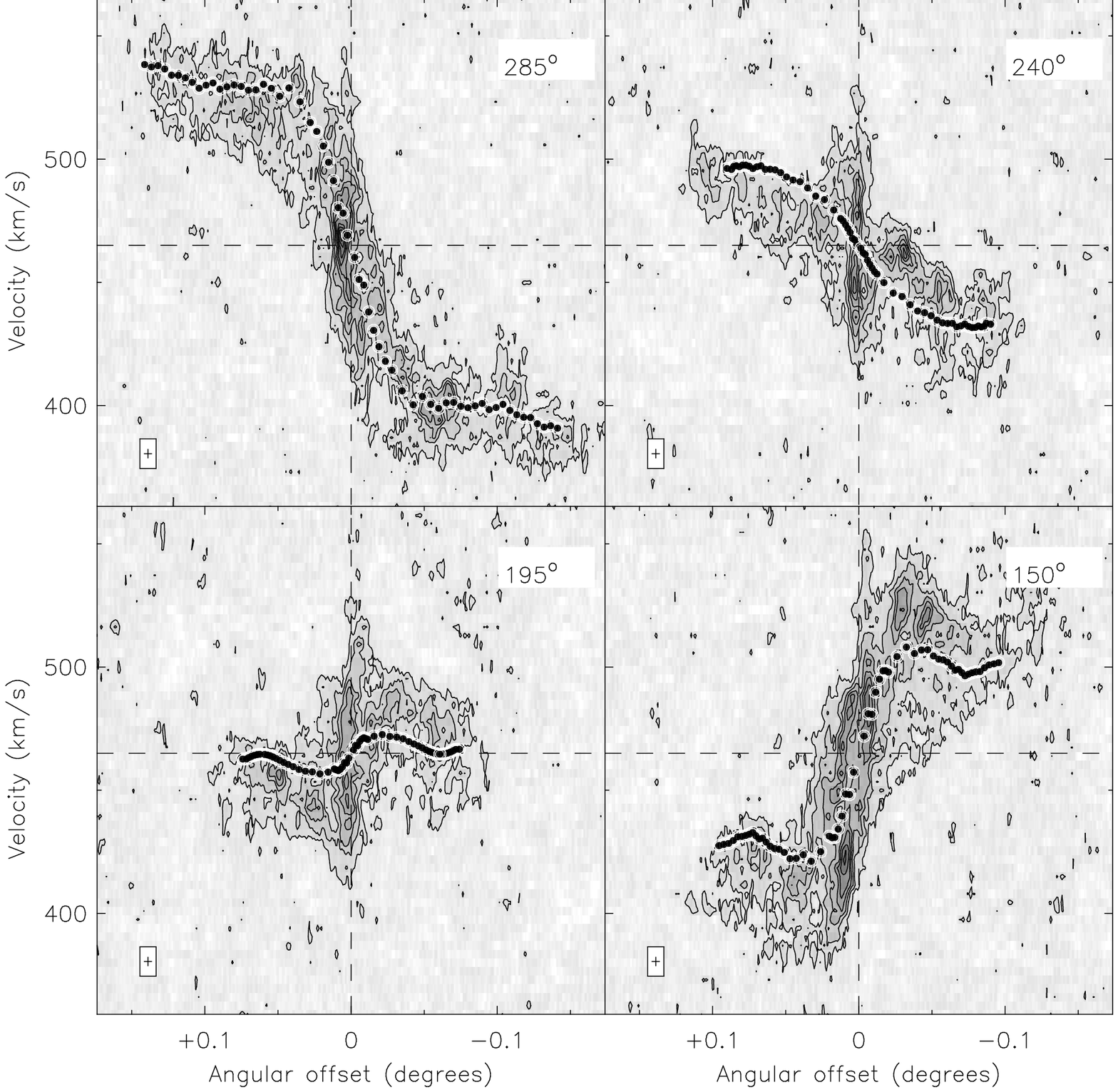}
	\caption{Position-velocity slices though the kinematic centre of NGC~2915 at various position angles.  Overlaid on each slice is the rotation curve of tilted ring model SI projected using the smoothed parameter radial profiles as shown in Fig. \ref{TR_results}.  The thickness of the position-velocity slices is that of a single pixel in the data cube (2.5$''$) and the velocity resolution is 3.5 km s$^{-1}$.  The dashed horizontal line shows the systemic velocity at 465 km s$^{-1}$ adopted for our kinematic analyses.  Contours start at 2$\sigma$ in steps of 2.5$\sigma$.}
	\label{pv_overlays}
	\end{centering}
\end{figure*}

For the final iteration, the $X_C$, $Y_C$ and $V_{sys}$ parameters were fixed to values that were constant with $r$ while the inclination and position angle radial profiles were smoothed and then fixed.   The smoothed versions of the profiles are shown as solid curves in panel A, B, C, D, E in Fig. \ref{TR_results} while the filled circles represent the parameter values just before they were smoothed and fixed.  Smoothing the $PA$ and $i$ radial distributions did not affect the final rotation curve in any noticeable manner.  As \citet{THINGS_deblok} point out, the smoothing only affects the point-to-point scatter of the profiles and in no way affects the resolution of the radial distributions.  To derive the final rotation curve, only $V_{rot}$ was allowed to vary.  One of the main sources of uncertainty in the derivation of the inner rotation curve is the rise in inclination of $\sim$~20\deg\ towards inner radii preferred by the tilted ring models.  It is not clear whether this rise is due to a genuine kinematic warp, or whether it is merely an artefact of the non-Gaussian line profiles near the galaxy's centre.  We therefore derived two final tilted ring models, one with and one without the rise in inclination (solid and dashed curves in panel D of Fig. \ref{TR_results} respectively).  Both models used the same $X_c$, Y$_c$, $V_{sys}$ and $PA$ radial profiles.  We hereafter refer to these two tilted ring models as models SI (Steep Inclination) and CI (Constant Inclination).  In panel F of Fig. \ref{TR_results}, the rotation curves of both models are presented.  In each of the panels A~-~E in Fig. \ref{TR_results}, the error bars represent the formal least squares errors from the ROTCUR fitting routine.  In panel F, the error bars represent the r.m.s. spread of the velocities in a single ring.

\subsection{Rotation curve}\label{TR_output}
The final tilted ring models fitted to the third-order Gauss-Hermite velocity field are presented in Fig. \ref{TR_results}.  While the stellar disk seen in the 3.6~\micron\ image (Fig.~\ref{IRAC_image}) extends as far as $R\sim 100''$, the inner portion of the rotation curve rises as $V(R)\propto R$ out to $R\sim 150''$.  Beyond this radius the rotation velocity remains almost constant.  For model SI, the average rotation velocity for $R\ge 187''$ is $V=81.9$~$\pm$~1.6~\kms.  Despite using different inclination profiles to derive the two final rotation curves, the maximum absolute difference between them is $V=9.3\pm12.2$~\kms\ at $R=136''$.  The average absolute difference for $R\le 187''$ is $V=4.3\pm 4.8$\kms.  The resulting two rotation curves are thus very similar to one another.  For each tilted ring model we constructed separate rotation curves for the approaching and receding sides of the galaxy.  In the case of model SI (Fig.~\ref{2vrots}), the rotation curve differs significantly between the two sides of the galaxy.  Within the steeply rising portion, differences of $\sim$~20~\kms\ are observed as well as at $R\sim 290''$.  The same is true in the case of model~CI.  \citet{swaters_lopsided} state that galaxies with asymmetric global profiles often have rotation curves that are more slowly rising on one side of the galaxy than the other.  This statement seems true in the case of NGC~2915.

To check that the two tilted ring models are consistent with the data, we used the smoothed radial parameter profiles to project the rotation curves onto various position-velocity slices extracted from the \hi\ data cube.  The overlays for tilted ring model SI are presented in Fig~\ref{pv_overlays}.  The projected rotation curves of the two tilted ring models are almost identical to one another with model SI fitting well the high-intensity regions of the position-velocity slices.  Given a set of fixed radial profiles, the ROTCUR routine will adjust the rotation curve so that the \emph{line-of-sight} velocities best match the data.  Thus although the circular rotation curves (i.e. panel F of Fig.~\ref{TR_results}) differ, the good agreement between the projected rotation curves is not surprising.

\begin{figure}
	\begin{centering}
 	\includegraphics[width=1\columnwidth]{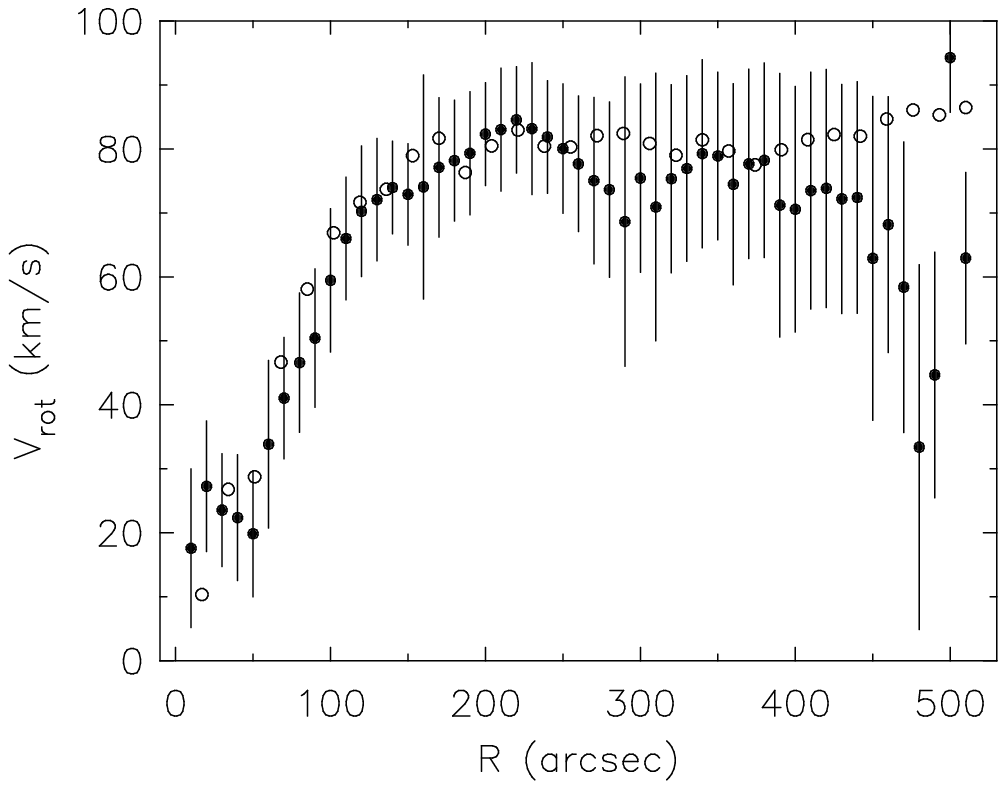}
	\caption{The model SI rotation curve (open circles) compared to the rotation curve derived by fitting a tilted ring model to the \hi\ velocity field extracted from the RW  \hi\ data cubes (black filled circles).   Error bars represent the r.m.s. spread of the velocities in a single ring of the tilited ring model fitted to the RW velocity field.  For the sake of clarity, error bars are not shown for the model SI rotation curve.}
	\label{10as_17as_vrots}
	\end{centering}
\end{figure}

Using the smoothed, fixed radial parameter profiles of model SI and ring widths of $dr=10''$, a tilted ring model was also fitted to the third-order Gauss-Hermite velocity field extracted from the RW \hi\ cube.  The resulting rotation curve is compared to the rotation curve of model SI in Fig.~\ref{10as_17as_vrots}.  The figure demonstrates that for $R\lesssim 250''$, the derived rotation curves are very similar to one another.  Beyond $R\sim 250''$ the filling factor of the RW \hi\ velocity field becomes too low for meaningful derivations of the rotation velocity.  For these reasons, only the rotation curves of tilted ring models CI and SI are used as mass model inputs in Sec.~\ref{mass_modeling}.

%-----------------------------------------------------------------------------------------------------------------------------------------------------------------
\section{Mass modeling}\label{mass_modeling}
%Their low stellar surface densities allow their rotation curves to be an almost direct measure of the distribution of DM.  They therefore allow us to test more directly the predictions of numerical cold dark matter simulations.
Dwarf galaxies such as NGC 2915 serve as useful probes of dark matter (DM).  In this section we derive two mass models for NGC 2915, one for each of the tilted ring models CI and SI.

The gravitational potential within a galaxy is determined by the combined gravitational potentials of all the mass components.  The derived 3.6~\micron\ surface brightness profile of NGC 2915 shows little deviation from an exponential light distribution at any radius.  We therefore concluded that the galaxy does not contain a significant central stellar bulge.  The total mass was treated as the sum of the masses of the stellar and gas disks as well as the DM halo.  Since $M\propto V^2$ at a given $R$, the rotation curves of the individual mass components, when summed in quadrature, will yield the square of total rotation curve, $V_{tot}$:
\begin{equation}
V_{tot}^2=\alpha_{gas} V_{gas}^2+\Upsilon_* V_*^2+V_{DM}^2,
\label{vtot} 
\end{equation}
where $V_{gas}$, $V_*$, and $V_{DM}$ are the contributions to the total rotation curve of the gas, the stars and the DM respectively.  $V_{tot}$ should match the observed rotation curve.  $\Upsilon_*$ is the stellar mass-to-light ratio used to convert the observed distribution of star light to a stellar mass distribution while $\alpha_{gas}=1.37$ is the scaling factor used for all mass models to take into account the contribution of helium to the rotation curve.
%------------------------------------------------------------------------------------------------------------------------------------------------------------
\subsection{Gas and stellar distributions}\label{bary_vrots}
We used the GIPSY task ROTMOD \citep{rotmod} to convert the observed mass distributions into rotation curves.  Each of the fitted tilted ring models (Sec. \ref{TR_output}) was used together with the \hi\ total intensity map to yield an \hi\ surface density profile.  These profiles were used as ROTMOD input which, assuming an infinitely thin gas disk, yielded two different \hi\ rotation curves.  Ellipses fitted to 3.6~\micron\ surface brightness isophotes suggest a constant inclination of $i\sim 55^{\circ}$ for the old stellar population.  Using this inclination and a position angle of  306\deg, a single surface density profile for this mass component was derived.  This surface density profile was extrapolated out to $R=510''$ to match the radial extent of the observed rotation curve, and then converted into a stellar rotation curve assuming a $\sec$h$^2$ vertical distribution \citep{van_der_kruit_1981} and $h/z_0=5$, yielding $z_0=0.12$~kpc.  

In principle the DM density distribution of a suitably rotating system can be determined knowing the contributions to the observed rotation curve from the gas and the stars.  The \hi\ rotation curve is directly related to the observed \hi\ distribution, assuming that the \hi\ is optically thin.  The situation is not as simple for the stellar component of a galaxy, however, since a stellar mass-to-light ratio is needed to link the stellar light and mass distributions.  Determining the stellar mass-to-light ratio from the rotation curve alone is problematic.  Given an observed rotation curve, equally well-fitted mass models can be obtained for a range of stellar mass-to-light ratios \citep[e.g.][]{van_albada_1985,swaters_thesis}.  This degeneracy therefore leads to uncertainties in stellar mass-to-light ratios determined solely by rotation curves.

\citet{bell_dejong} used spectroscopic and photometric spiral galaxy evolution models to show that a reasonable correlation exists between the optical and infrared stellar mass-to-light ratios and the colours of integrated stellar populations.  They further showed that the relative trends between model stellar mass-to-light ratios and colours are fairly insensitive to uncertainties in stellar population and galaxy evolution modeling.  \citet{THINGS_oh} used stellar population synthesis models (together with a scaled Salpeter 1955 IMF) to determine a relationship between the stellar mass-to-light ratios in the 3.6~\micron\ and $K$ bands ($\Upsilon_*^{3.6}$ and $\Upsilon_*^K$  respectively):
\begin{equation}
\Upsilon_*^{3.6}=0.92\times \Upsilon_*^K-0.05.
\label{emp_ML}
\end{equation}
From \citet{bell_dejong}, the relation between the $B-V$ colours and the $K$-band mass-to-light ratio is
\begin{equation}
\log_{10}\left(\Upsilon_*^K\right)=b^K\times (B-V)+a^k,
\label{k-colour}
\end{equation}
where $a^K$ and $b^k$ for varying colours and metallicities are provided in their Table 4.  By combining these two equations we can estimate the stellar mass-to-light ratio at 3.6~\micron.  We use $a^K=-0.59$ and $b^K=0.60$ for $B-V$ colours as used by \citet{THINGS_oh} for IC 2574, the justification being that IC~2574 is also a low metallicity, star-forming dwarf galaxy with $B-V\sim 0.6$, much like NGC 2915.  $B-V\sim 0.3 \pm 0.2$ for NGC~2915 \citep{RC3,schlegel} thereby yielding (from Eqn.~\ref{k-colour}) $\Upsilon_*^K\approx 0.39$.  Substituting this value into Eqn. \ref{emp_ML} yields $\Upsilon_*^{3.6}\approx 0.3$.  \citet{leroy_THINGS}, applying the \citet{bell_et_al_2003} relation between $B-V$ colour and $\Upsilon_*^K$ and assuming a \cite{kroupa_IMF} IMF, found $\Upsilon_*^K=~0.48-0.60$ for the $K-$band stellar mass-to-light ratios of all of the THINGS galaxies in their sample.

%---------------------------------------------------------------------------------------------------------------------
\subsection{Dark matter halos}
The way in which dark matter is distributed on galactic length-scales has been a topic of much debate.  Over the past decade, the constant increase in computational power has allowed detailed $N$-body simulations of the clustering properties of cold dark matter particles to be carried out.  One of the first large-scale studies was carried out by \citet{NFW} in which they studied the equilibrium density profiles of dark matter halos in a hierarchial galaxy clustering scenario.  Their results suggested the existence of a ``universal dark matter density profile'' \citep{NFW_1997} which is independent of clustering scale, mass or size as well as the power spectrum of initial fluctuations.  Their parameterisation of this universal profile is
\begin{equation}
{\rho(r)_{NFW}\over \rho_{crit}}= {\delta_c \over {(r/r_s)(1+r/r_s)^2}}
\end{equation}
where $\delta_c$ is a measure of the density of the Universe at the time of collapse of the DM halo, $r_s$ is the characteristic scale radius and $\rho_{crit}=3H^2/8\pi G$ is the critical density required for closure.  This profile scales as $r^{\alpha}$, with $\alpha=~-3$ and $\alpha=~-1$ at large and small radii respectively, thereby predicting extremely steep inner density profiles of DM halos, known as central cusps.  

The Aquarius Project \citep{aquarius_subhalos} recently completed a set of high-resolution numerical simulations to study the formation and structure of galaxy-sized DM halos in the standard $\Lambda$CDM cosmology.  \citet{navarro_2008} reported that the spherically averaged density profile of the six different-sized simulated galaxy halos becomes progressively shallower inwards and, at the innermost resolved radius, has a logarithmic slope $\alpha = \mathrm{d}\ln \rho/\mathrm{d}\ln r \gtrsim -1$.   These predictions of cuspy inner DM density profiles can be tested against real, dark-matter-dominated galaxies.  If cuspy NFW halos are found, constraints can be put on halo concentrations and cosmological parameters.  The rotation curve resulting from an NFW density profile is
\begin{equation}
\left({V_{NFW}\over V_{200}}\right)^2={\ln (1+cx)-cx/(1+cx)\over x\ln(1+cx)-c/(1+c)},
\end{equation}
where $V_{200}$ is the circular rotation speed at $r_{200}$, the radius at which the density of the DM halo equals 200 times the critical density, $\rho_{crit}$; $x=~r/r_{200}$ is the radius in units of the virial radius and $c=r_{200}/r_s$ is the ``concentration'' parameter of the halo.  

Low surface brightness dwarf galaxies are thought to be dark-matter-dominated down to small radii \citep{deblok_mcgaugh_1996} and can therefore be more directly compared to the cold dark matter simulations.  When this is done, it is often the case that the density profiles of simulated DM halos are too steep to fit the rotation curves of dwarf galaxies \citep{moore_1994,flores_primack,deblok_mcgaugh_1996,marchesini_2002}.  The distribution of the DM in these dwarfs is instead consistent with approximately constant inner dark matter densities.  A particular dark matter density profile that describes the DM distribution of dwarf systems well is that of the pseudo-isothermal sphere \citep{galactic_dynamics}:
\begin{equation}
\rho(r)_{ISO}=\rho_0\left(1+\left({r\over r_c}\right)^2\right)^{-1},
\end{equation}
where $\rho_0$ is the central DM density and $r_c$ is the core radius.  This parameterisation, which has no particular physical justification, has a constant density core.  The rotation curve resulting from such a density profile is
\begin{equation}
V_{ISO}=\left(4\pi G \rho_0 r_c^2\left[1-{r_c\over r}\arctan\left(r\over r_c\right)\right]\right)^{1/2}.
\end{equation}
%-------------------------------------------------------------------------------------------------------------------------------------------------------
\subsection{Fitted models}
We used the GIPSY task ROTMAS to construct mass models for NGC~2915.  The task subtracts from the observed rotation curve the scaled rotation curves of the stellar and gas disks.  It then fits to the residuals a rotation curve corresponding to a particular parameterisation of the DM halo.  When matching the total and observed rotation curves, the points were uniformly weighted to ensure that each point in the rotation curve contributed equally to the fit.  This essentially reduced the fitting procedure to a non-weighted least-squares fitting method.

As previously mentioned, each of the rotation curves from the two tilted ring models fitted to the \hi\ velocity field were used as mass modeling input.  For each of these mass models we carried out fits for a pseudo-isothermal sphere and an NFW halo.  Upper and lower limits for the stellar mass-to-light ratio were determined by scaling the contribution of the stellar rotation curve to the total rotation curve.  Under the so-called ``maximum disk assumption'', the stellar rotation curve is scaled up to contribute maximally to the observed rotation curve at inner radii.  Such a scaling usually allows the stellar disk to explain most of the inner rotation curve \citep{van_albada_1985,swaters_thesis}.  Under the ``minimum disk assumption'', the contribution of the stellar disk is made as low as possible (often zero) while still allowing a good fit of the total rotation curve to the observed rotation curve.  Each of the halo fits was divided into 3 sub-cases: A fit 1) with $\Upsilon_*$ fixed to our predetermined value of 0.3; 2) using a maximum disk assumption; and 3) using a minimum disk assumption.  For the minimum disk cases, $\Upsilon_*$ was fixed to zero (no contribution from the stellar disk to the observed rotation curve).  For the maximum disk cases, regardless of the DM halo parameterisation, $\Upsilon_*$ was fixed to 0.64 and 0.44 for mass models SI and CI respectively.  In all models, $\alpha_{gas}$ was fixed to 1.37. Thus, six mass models were fitted to each of the 2 rotation curves derived from the tilted ring modeling.  %Table \ref{mass_modeling_results} lists the various $\Upsilon_*$ scaling factors used for the various mass models.
%---------------------------------------------------------------------------------------------------------------------------------------------------------
\subsection{Results}
\begin{table*}
\begin{center}
% use packages: array
\caption{Mass modeling results}
\label{mass_modeling_results}
\scriptsize{
\begin{tabular}{cccccccc}
\hline
\hline
\\
&&1 & 2 & 3 & 4 & 5 &6\\ 
\\
&DM halo  & $ISO$ & $ISO_{max}$  & $ISO_{min}$  & $NFW$  & $NFW_{max}$ &$NFW_{min}$\\ 
\\
\hline
\\
&\textbf{Model SI}\\
\\
1&$\chi^2_{red}$ & 38.1 & 59.8 & 28.2 & 62.5 & 93.4 & 44.7 \\

2&r.m.s. (km s$^{-1}$) & 0.56 & 1.30 & 0.79 & 0.75 & 1.37 & 0.40\\
%$\alpha$ & 1.37 & 1.37 & 1.37 & 1.37 & 1.37 & 1.37\\

3& $\Upsilon_*$ ($M_{\odot}/L_{\odot}$)& 0.3 & 0.64 & 0.0 & 0.3 & 0.64 & 0.0\\

4& $\rho_0$ (10$^{-3}$ $M_{\odot}$ pc$^{-3}$) & 100.6 $\pm$ 23.3 & 56.9 $\pm$ 17.8 & 178.7 $\pm$ 33.0 & ... & ... & ...\\

5& $R_c$ (kpc) & 1.3 $\pm$ 0.2 & 1.7 $\pm$ 0.4 & 0.9 $\pm$ 0.1 & ... & ... & ...\\

6& $V_{200}$ (km s$^{-1}$) & ... & ... & ... & 69.6 $\pm$ 7.9 & 83.5 $\pm$ 20.1 & 62.7 $\pm$ 4.1\\

7& $C$ & ... & ... & ... & 11.7 $\pm$ 2.2 & 7.9 $\pm$ 2.7 & 15.6 $\pm$ 2.0\\
\\
&At last measured pt:\\
\\
8& $M_{tot}$ (10$^{8}$ \msun) & 151.4 & 151.3 & 151.6 & 151.1 & 154.2 & 148.3\\

9& $M_{HI}/M_{tot}$ (10$^{-3}$) & 43.84 & 43.85 & 43.79 & 43.92 & 43.03 & 44.76\\

10& $M_{DM}/M_{tot}$ (10$^{-2}$) & 92.94 & 89.91 & 95.62 & 92.93 & 90.10 & 95.52\\

11& $M_{DM}/M_{lum}$ & 6.68 & 4.47 & 21.84 & 6.67 & 4.57 & 21.34\\

12& $M_{tot}/L_B$ ($M_{\odot}/L_{\odot}$) & 140.72 & 140.69 & 140.91 & 140.47 & 143.37 & 137.9\\
\\
\hline
\\
& \textbf{Model CI}\\
\\
1& $\chi^2_{red}$ & 46.7 & 53.6 & 37.5 & 78.1 & 89.1 & 61.2\\

2& r.m.s. (km s$^{-1}$) & 0.74 & 1.00 & 0.27 & 1.03 & 1.28 & 0.66\\

%$\alpha$ & 1.37 & 1.37 & 1.37 & 1.37 & 1.37 & 1.37\\

3& $\Upsilon_*$ ($M_{\odot}/L_{\odot}$)& 0.3 & 0.44 & 0.0 & 0.3 & 0.44 & 0.0\\

4& $\rho_0$ (10$^{-3}$ $M_{\odot}$ pc$^{-3}$) & 127.9 $\pm$ 34.9 & 103.2 $\pm$ 31.1 & 227.6 $\pm$ 50.1 & ... & ... & ...\\

5& $R_c$ (kpc) & 1.10 $\pm$ 0.2 & 1.2 $\pm$ 0.2 & 0.8 $\pm$ 0.1 & ... & ... & ...\\

6& $V_{200}$ (km s$^{-1}$) & ... & ... & ... & 66.4 $\pm$ 8.3 & 70.3 $\pm$ 11.5 & 60.7 $\pm$ 4.7\\

7& $C$ & ... & ... & ... & 12.9 $\pm$ 2.8 & 11.2 $\pm$ 2.9 & 17.0 $\pm$ 2.5\\
\\
& At last measured pt:\\
\\
8& $M_{tot}$ (10$^{8}$ \msun) & 138.8 & 138.3 139.8 & 137.6 & 138.4 136.3\\
 
9& $M_{HI}/M_{tot}$ (10$^{-3}$) & 47.29 & 47.43 & 46.92 & 47.66 & 47.42 & 48.16\\

10& $M_{DM}/M_{tot}$ (10$^{-2}$) & 92.34 & 91.13 & 95.31 & 92.28 & 90.92 & 95.18\\

11& $M_{DM}/M_{lum}$ & 6.11 & 5.15 & 20.31 & 6.06 & 5.01 & 19.77\\

12& $M_{tot}/L_B$ ($M_{\odot}/L_{\odot}$) & 129.00 & 128.62 & 130.01 & 127.99 & 128.64 & 126.68\\
 \end{tabular}}
\end{center}
\begin{flushleft}
\textbf{Comments on rows:} Row~1:~Reduced $\chi^2$ goodness-of-fit statistic; Row~2:~r.m.s. of the difference between the observed and total rotation curves; Row~3:~stellar mass-to-light ratio; Row~4:~central density for pseudo-isothermal sphere; Row~5:~core radius for pseudo-isothermal sphere, Row~6:~circular rotation speed at virial radius for NFW halo; Row~7:~concentration parameter for NFW halo; Row~8:~dynamical mass; Row~9:~\hi\ to total mass ratio; Row~10:~Dark to total mass ratio; Row~11:~Dark to luminous mass ratio; Row~12:~Total mass to $B$-band light ratio.\\
\textbf{Comments on columns:} For each model, columns 1 and 4 show the mass modelling results for the case in which $\Upsilon_*$ is fixed to the predetermined value of 0.3.  Columns 2 and 5 show the results for the maximum disk case while columns 3 and 6 are for the minimum disk case.
\end{flushleft}
\end{table*}

\begin{figure*}
	\begin{centering}
	\includegraphics[angle=0,width=2\columnwidth]{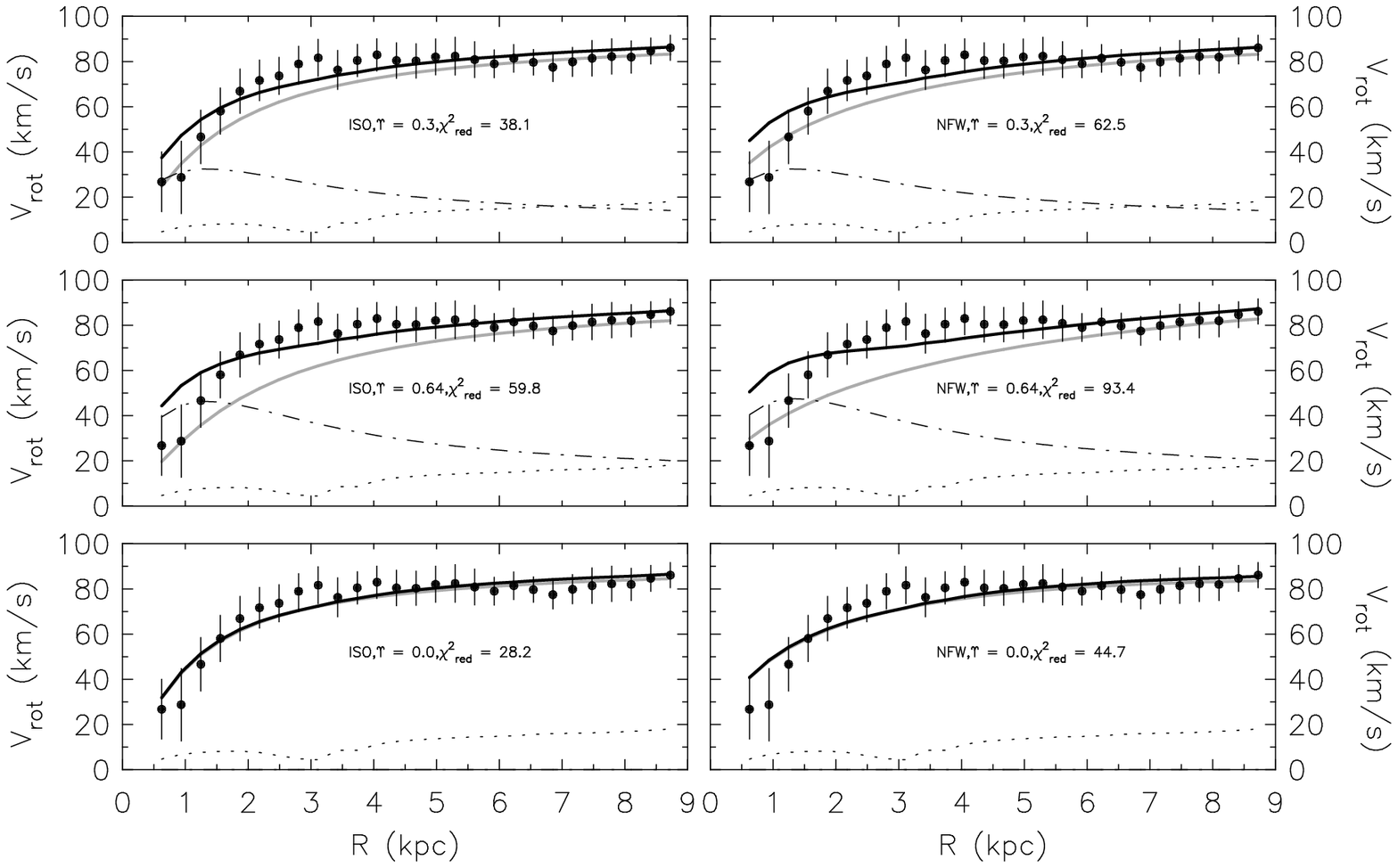}
	\caption{Pseudo-isothermal sphere (ISO) and NFW rotation curve fits for NGC 2915.  The panels on the left show the fits using the pseudo-isothermal sphere while the panels on the right show the fits using an NFW halo.  The top panels show the fits when the stellar rotation curve is scaled by our predetermined value of $\Upsilon_*=0.3$ while the middle and bottom panels shows the fits for the maximum and minimum disk cases respectively.  In all panels the black filled circles represent the observed rotation curve while the error bars represent the r.m.s. spread of velocities in a given ring.  The black dot-dash curve shows the rotation curve of the gas while the black dotted curve shows the rotation curve of the stellar disk.  The solid grey curve shows the resulting rotation curve of the DM halo.  Finally, the solid black curve represents the total rotation curve resulting from the best-fit model.  In each panel the $\chi^2$ goodness-of-fit statistic is presented.  The parameter radial profiles used to constructed the stellar and gas rotation curves were those of tilted ring model SI.  The observed rotation curve is also from tilted ring model SI.}
	\label{rotmas_TR_A_plot}
	\end{centering}
\end{figure*}

The fitted parameters for all of the mass models are summarised in Table \ref{mass_modeling_results}.  The results for mass model SI are also presented in Fig. \ref{rotmas_TR_A_plot}.  This figure shows that NGC 2915 is dark-matter-dominated at nearly all radii (depending on modeling assumptions).  We calculate mass ratios as well as the total mass to $B$-band light ratio at the last measured point of each fitted total rotation curve.  For mass models SI and CI, DM  constitutes, on average, 92.8$\%$ of the dynamical mass. This dark-matter-dominance is confirmed by the total-mass-to-light ratio of NGC 2915 which is 141 \msun/$L_{\odot}$ and 129 \msun/$L_{\odot}$ on average for mass models SI and CI respectively.  These estimates are almost double that of the 76~\msun/$L_{\odot}$ upper limit set by \citet{meurer2} thereby making NGC~2915 one of the most dark-matter-dominated late-type dwarf galaxies known.

The second main mass-modeling result that Fig. \ref{rotmas_TR_A_plot} demonstrates is that, for either parameterisation of the DM halo, the steeply rising portion of the observed rotation curve is poorly matched by the total rotation curve.  It is usually the stellar rotation curve that is used to match the inner observed rotation curve \citep{swaters_thesis}, yet no such scaling is useful in the case of NGC 2915 due to the intrinsically different shapes of the stellar and observed rotation curves.  The stellar disk is clearly contained well within the steeply rising portion of the observed rotation curve which continues to rise as $V(R)\propto R$ out to $R\sim 150''$.  The best-fitting mass models are those in which the stellar rotation curve contributes zero to the observed total rotation curve.  Figure \ref{rotmas_TR_A_plot} shows that when the fitting routine attempts to fit the inner observed rotation curve with the DM rotation curve, the result is an over-estimation of the observed rotation velocities for $R\lesssim 2$ kpc (110$''$) and an under-estimation for radii 2~kpc~$\lesssim R \lesssim$~5.5~kpc. The poor matches between the total and observed inner rotation curves do not allow a particular DM halo parameterisation to be confidently ruled out or confirmed.  Furthermore, the observed non-Gaussian line profiles near the centre of NGC 2915 (Sec. \ref{line_profiles}) lead to uncertainties in the shape of the inner portion of the observed rotation curve.  These, in turn, lead to associated uncertainties in the shapes of the inner portions of the fitted DM halos.  Assuming that the stars are less susceptible than the gas to the effects of non-cricular motions, spectroscopic observations of the stars would have to be carried out to better constrain the inner rotation curve. 

What is the explanation for the discrepancy between the observed and total rotation curves?  Have we failed to include a significant mass component during the decomposition process?  A stellar bulge cannot be the answer since our derived 3.6~\micron\ surface brightness profile shows no significant deviation from an exponential light distribution.  A significant mass of molecular gas at inner radii is a possibility, yet CO emission in low-mass galaxies is usually weak or not detected \citep{taylor_1998,leroy_2005}.  However, the conversion factor used to convert CO luminosity to molecular gas mass is largely uncertain.%  A possible central compact object such as an intermediate-mass black hole is no solution either since is would be contained well within the knee.%  Assuming a dust temperature of 30 K\footnote{} together with the 100~\micron\ IRAS flux of ... from \citet{qwe}, we estimate\footnote{Using the equation from Sec. 3.3 of \citet{qwe}} an upper limit for the dust mass within stellar disk of NGC 2915 to be 1.44 $\times$ 10$^4$ \msun

If we consider the mass-modeling results as they appear in Table \ref{mass_modeling_results} then the pseudo-isothermal sphere allows for the best match to the observed rotation curve.  The fitted pseudo-isothermal sphere, under the minimum disk assumption, has a core density and a core radius of $\rho_0=0.17\pm~0.03$~$M_{\odot}$~pc$^{-3}$ and $r_c=0.9\pm 0.1$~kpc for model SI.  These results are similar to those of \citet{meurer2} who, for their favoured mass model, estimated $\rho_0=0.10\pm~0.02$ ~$M_{\odot}$~pc$^{-3}$ and $r_c=1.23\pm 0.15$~kpc. We also fitted the mass models with $\Upsilon_*$ as a free parameter yet always found the best-fitting value to be negative, consistent with the mass modeling results of \citet{meurer2}.

\begin{figure}
	\begin{centering}
	\includegraphics[angle=0,width=1\columnwidth]{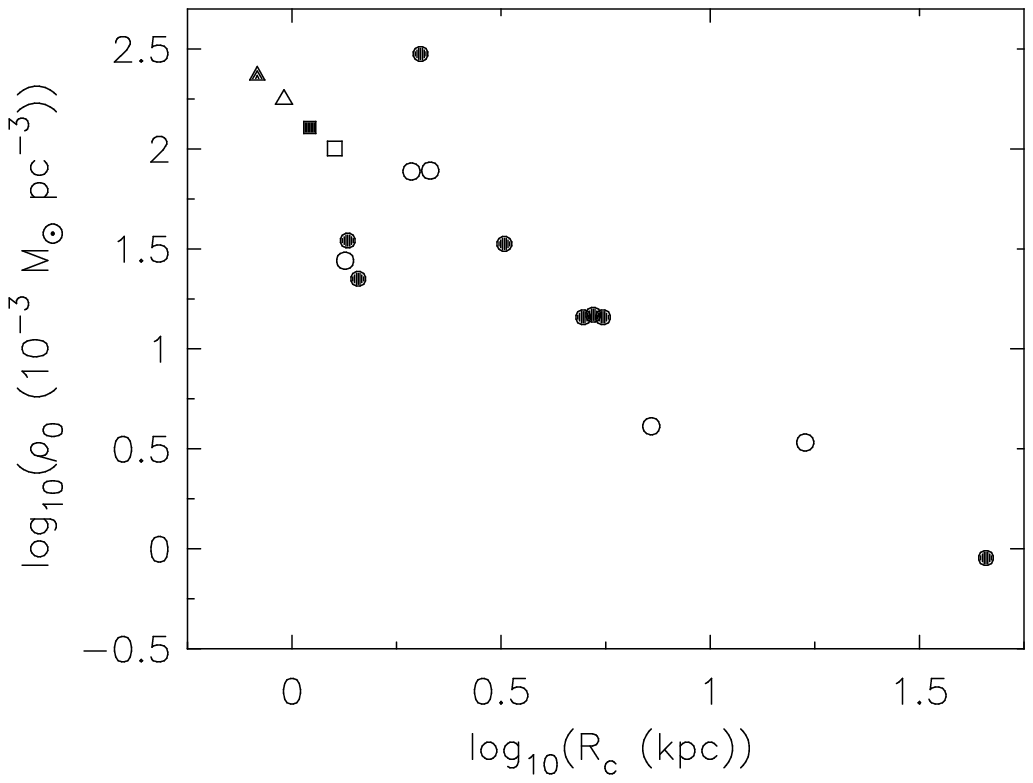}
	\caption{Mass modeling results for the minimum disk and $\Upsilon_*=0.3$ cases in which the DM halo is modelled as a pseudo-isothermal sphere, compared to the mass modeling results of the THINGS sample of \citet{THINGS_deblok}.  The black-filled and open triangles correspond to the minimum disk results for models SI and CI respectively while the black-filled and open squares correspond to the $\Upsilon_*=0.3$ results for models SI and CI respectively.  The mass modeling results from THINGS are shown as open circles for dwarf galaxies and filled circles for all other galaxies.}
	\label{THINGSmm}
	\end{centering}
\end{figure}

In Fig. \ref{THINGSmm} we compare our derived pseudo-isothermal halo parameters for NGC 2915 to those derived for the THINGS sample from \citet{THINGS_deblok}.  The infrared stellar mass-to-light ratios used in their mass models were derived from the 3.6~\micron\ images in combination with stellar population synthesis arguments.  Our models for the minimum disk and $\Upsilon_*=0.3$ cases for both models SI and CI are shown in Fig. \ref{THINGSmm}.  This comparison suggests that NGC 2915 has a central DM core that is very compact and dense compared to other late-type systems.  

%Finally, it remains to be said that we have not taken into account the possible effects of the central non-circular gas motions on the inferred DM halo properties.  We also have not considered the possibility of a tri-axil DM halo.  These are two types of systematic effects that can make intrinsically cuspy profiles appear cored.  Quantifying these sorts of effects on the inferred DM halo properties will not 

%plot made using /Users/edelson/2915_paper1/mass_modeling/THINGS_mass_modeling_plot.py

%-----------------------------------------------------------------------------------------------------------------------------------------------------------------
\section{Summary and conclusions}\label{conclusions}
We have obtained new deep, high-resolution \hi\ synthesis observations of NGC~2915 and have used them to carry a detailed study of the \hi\ distribution and dynamics of this galaxy.  The galaxy has very different optical and \hi\ properties.  Optically, it is similar to other dwarf galaxies, consisting of a small stellar disk.   Surrounding the stellar disk, however, is a huge \hi\ disk extending out to galactocentric radii greater than $R= 510''$ with well-defined spiral structure in its outer parts.  We estimated the total mass of the \hi\ disk to be $M_{HI}\sim 4.4 \times 10^8$ \msun.

Our high resolution, high sensitivity observations clearly resolve the inner \hi\ distribution of NGC~2915 into two \hi\ concentrations separated by $\sim$ 1.1 kpc.  This observation discredits the claim by previous investigators that the central gas disk of the system is made up of a massive \hi\ bar.   We have shown the shapes of \hi\ line profiles to be significantly non-Gaussian in the vicinity of these central \hi\ over-densities, being split by $\sim$~30~\kms\ on average over the spatial extent of the lower \hi\ concentration.  These profiles are indicative of non-circular velocity components within the gas near the centre of NGC~2915.  In our forthcoming paper (Elson et al., 2010, in prep.) we investigate in further detail the central gas dynamics of this system.  Linked to these non-circular velocity components, we have provided speculative evidence for the possibility of cold gas accretion from the inter-galactic medium onto the outer disk of NGC~2915 which, if confirmed, would have significant implications for the evolutionary history of this system.  These complicated dynamics lead to the new picture that NGC~2915 is not a simple, isolated, low-surface-brightness galaxy but rather an evolving system with a complex interplay between its various mass components.

The main focus of this paper has been the regular \hi\ dynamics of NGC~2915.  We have  fitted a tilted ring model to the \hi\ velocity field.  The best-fitting model was one in which the \hi\ disk is severely warped, with the inner and outer disks inclined at $i\sim$ 70 - 75 \deg\ and $i \sim$ 55\deg\ respectively.  This result is, however, made uncertain by the non-Gaussian line profiles near the centre of the galaxy.  We therefore also fitted a model with an almost constant inclination of $i\sim$~55\deg\ for the entire \hi\ disk.  Both models yielded a rotation curve typical of a late-type spiral with $V(R)\propto R$ rotation out to $R\sim 150''$ and constant velocity thereafter.  Rotation curves separately derived for the approaching and receding sides of the galaxy differed significantly within their steeply rising portions, providing the clear evidence that NGC~2915 is a kinematically lopsided galaxy.

Our results show that these rotation curves cannot be accounted for by the stellar or gas disks of NGC~2915.  By using the observed rotation curves as mass model inputs, we found the galaxy to be dark-matter-dominated at nearly all radii with the stellar disk unable to account for the $V(R)\propto R$ portion of the rotation curve.  The best-fitting mass model is one in which the stellar disk does not contribute to the observed rotation curve and in which the inferred dark matter halo is parameterised as a pseudo-isothermal sphere.  This model has a central core density of $\rho_0= 0.17 \pm 0.03$~\msun~pc$^{-3}$ and a core radius of $R_c = 0.9 \pm 0.1$~kpc.  The non-circular gas motions at inner radii lead to uncertainties in the fitted dark matter halo parameters.  These, together with the fact that neither DM halo parameterisation allows for an accurate match of the observed rotation curve at inner radii, do not allow a particular DM halo parameterisation to be confidently ruled out or confirmed.  Further observations and modeling are required to more accurately determine the inner rotation curve of NGC~2915 and to therefore more confidently discriminate between various parameterisations of the DM halo.  It is clear, however, that NGC~2915 has a very dense and compact dark matter core.  At the last measured point on the rotation curve, this galaxy has a total-mass-to-light ratio of $M/L\sim$~140~\msun/L$_{\odot}$, almost twice that of the upper limit placed my \citet{meurer2}, thereby making it one of the most dark-matter-dominated galaxies known.

\section{Acknowledgments}
The work of ECE is based upon research generously supported by the South African SKA project.  ECE would like to thank Prof. Renzo Sancisi for his enthusiasm and endless patience in discussing and helping to analyse the \hi\ data.  All authors acknowledge funding recieved from the South African National Research Foundation.  The work of WJGDB is based upon research supported by the South African Research Chairs Initiative of the Department of Science and Technology and the National Research Foundation.  The Australian Telescope Compact Array is part of the Australia Telescope which is funded by the Commonwealth of Australia for operation as a National Facility managed by CSIRO.  This work is based [in part] on observations made with the \emph{Spitzer} Space Telescope, which is operated by the Jet Propulsion Laboratory, California Institute of Technology under a contract with NASA.  Finally, all authors thank the referee, Claude Carignan, for constructive comments that improved the quality of the paper.
%-----------------------------------------------------------------------------------------------------------------------------------------------------------------

%\bibliographystyle{mn2e} 
%\bibliography{/Users/edelson/latex/bibliography} % 

\label{lastpage}

\end{document}